\pdfoutput=1

\documentclass[11pt,twoside,a4paper,cmspaper,final,collab]{cms-tdr}
\def\svnVersion{367042}\def\cmsCernNoTag{CERN-EP/2016-073}\def\cmsCernDate{\today}\def\cmsMessage{Published in the Journal of High Energy Physics as \href{http://dx.doi.org/10.1007/JHEP08(2016)119}{\doi{10.1007/JHEP08(2016)119.}}}

\begin{document}\cmsNoteHeader{FSQ-13-008}

\hyphenation{had-ron-i-za-tion}
\hyphenation{cal-or-i-me-ter}
\hyphenation{de-vices}
\RCS$Revision: 364778 $
\RCS$HeadURL: svn+ssh://svn.cern.ch/reps/tdr2/papers/FSQ-13-008/trunk/FSQ-13-008.tex $
\RCS$Id: FSQ-13-008.tex 364778 2016-08-17 08:37:33Z jjhollar $
\newlength\cmsFigWidth
\ifthenelse{\boolean{cms@external}}{\setlength\cmsFigWidth{0.85\columnwidth}}{\setlength\cmsFigWidth{0.4\textwidth}}
\cmsNoteHeader{FSQ-13-008}
\title{Evidence for exclusive $\PGg\PGg \to \PWp \PWm$ production and constraints on anomalous quartic gauge couplings in $\Pp\Pp$ collisions at $\sqrt{s}=7$ and 8\TeV}

\providecommand{\PHANTOM}{\textsc{phantom}\xspace}
\renewcommand{\Pl}{\ensuremath{\ell}\xspace}
\providecommand{\Plp}{\ensuremath{\ell^+}\xspace}
\providecommand{\Plm}{\ensuremath{\ell^-}\xspace}

\date{\today}

\abstract{
A search for exclusive or quasi-exclusive $\PGg\PGg\to\PWp\PWm$ production, via $\Pp\Pp \to \Pp^{(*)}\PWp\PWm\Pp^{(*)}\to \Pp^{(*)}\Pgm^{\pm}\Pe^{\mp}\Pp^{(*)}$ at
$\sqrt{s}=8$\TeV, is reported using data corresponding to an integrated luminosity of 19.7\fbinv. Events are selected by requiring the presence of an electron-muon pair with large transverse momentum $\pt(\PGm^{\pm}\Pe^{\mp})>30$\GeV, and no associated
charged particles detected from the same vertex. The 8\TeV results are combined with the previous 7\TeV results (obtained for 5.05\fbinv of data).
In the signal region, 13 (2) events are observed over an expected background of $3.9 \pm 0.6$ ($0.84 \pm 0.15$) events for 8 (7)\TeV, resulting in a combined excess of $3.4 \sigma$ over the background-only hypothesis.
The observed yields and kinematic distributions are compatible with the standard model prediction for exclusive and quasi-exclusive $\PGg\PGg \to \PWp\PWm$ production.
Upper limits on the anomalous quartic gauge coupling operators $a^{\PW}_{0,C}$ (dimension-6) and $f_{M0,1,2,3}$ (dimension-8),
the most stringent to date, are derived from the measured dilepton transverse momentum spectrum.}

\hypersetup{%
pdfauthor={CMS Collaboration},%
pdftitle={Evidence for exclusive gamma-gamma to W+ W- production and constraints on anomalous quartic gauge couplings at sqrt(s) = 7 and 8 TeV},%
pdfsubject={CMS},%
pdfkeywords={CMS, physics, forward physics, exclusive processes, gamma-gamma interaction, anomalous quartic gauge couplings}}

\maketitle
\section{Introduction}

\begin{figure}[htb]
\centering
\includegraphics[width=.33\textwidth]{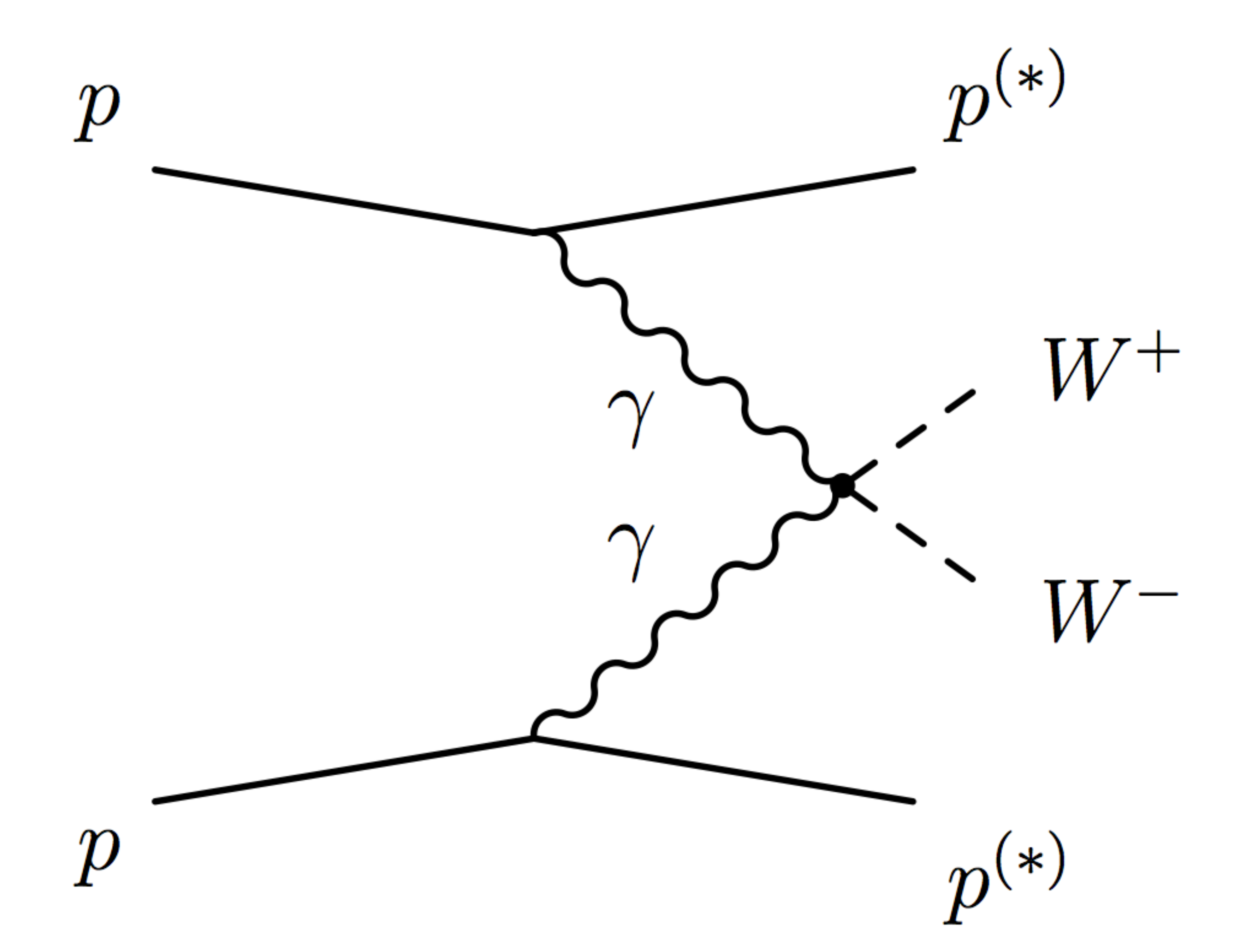}\hfill
\includegraphics[width=.33\textwidth]{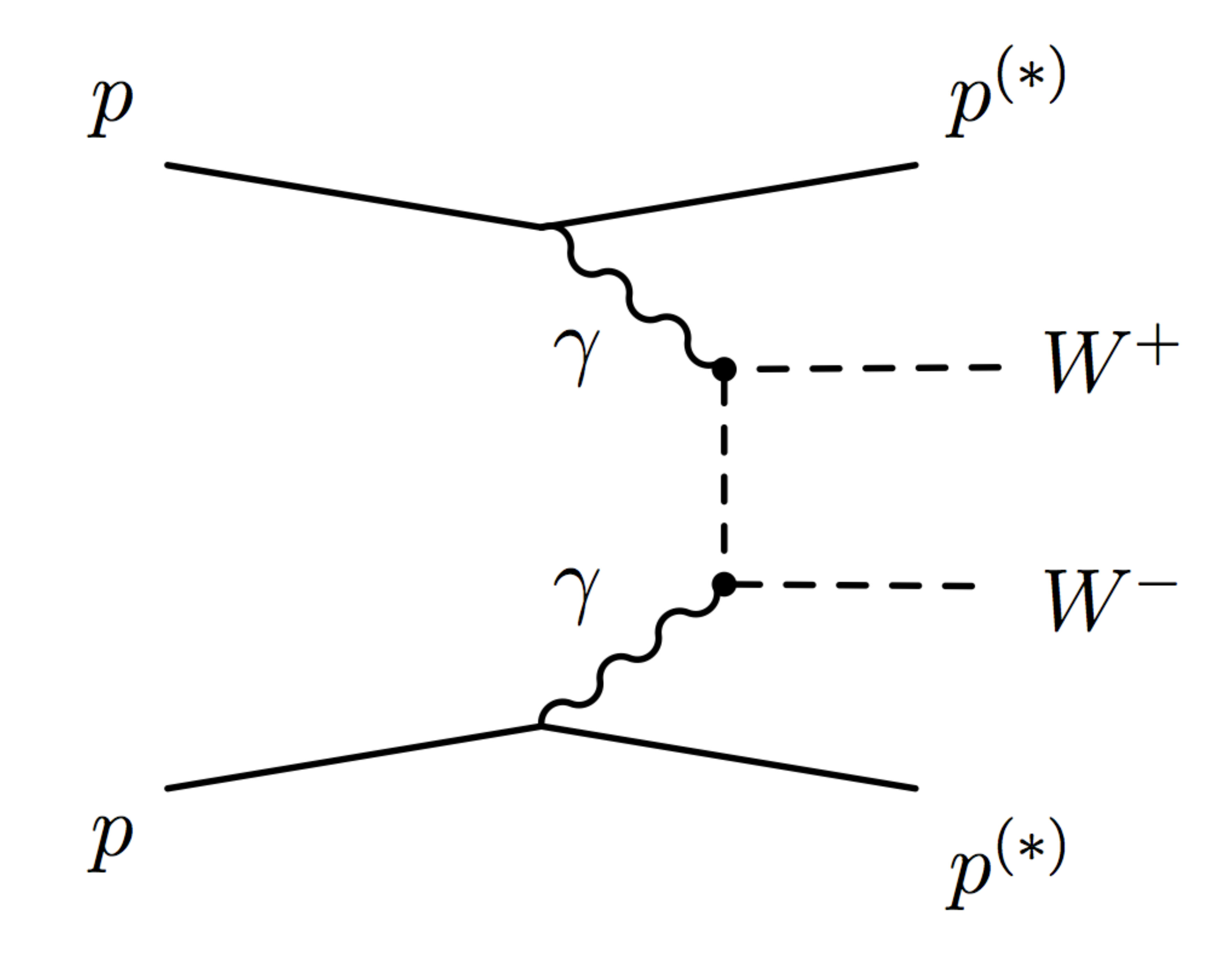}\hfill
\includegraphics[width=.33\textwidth]{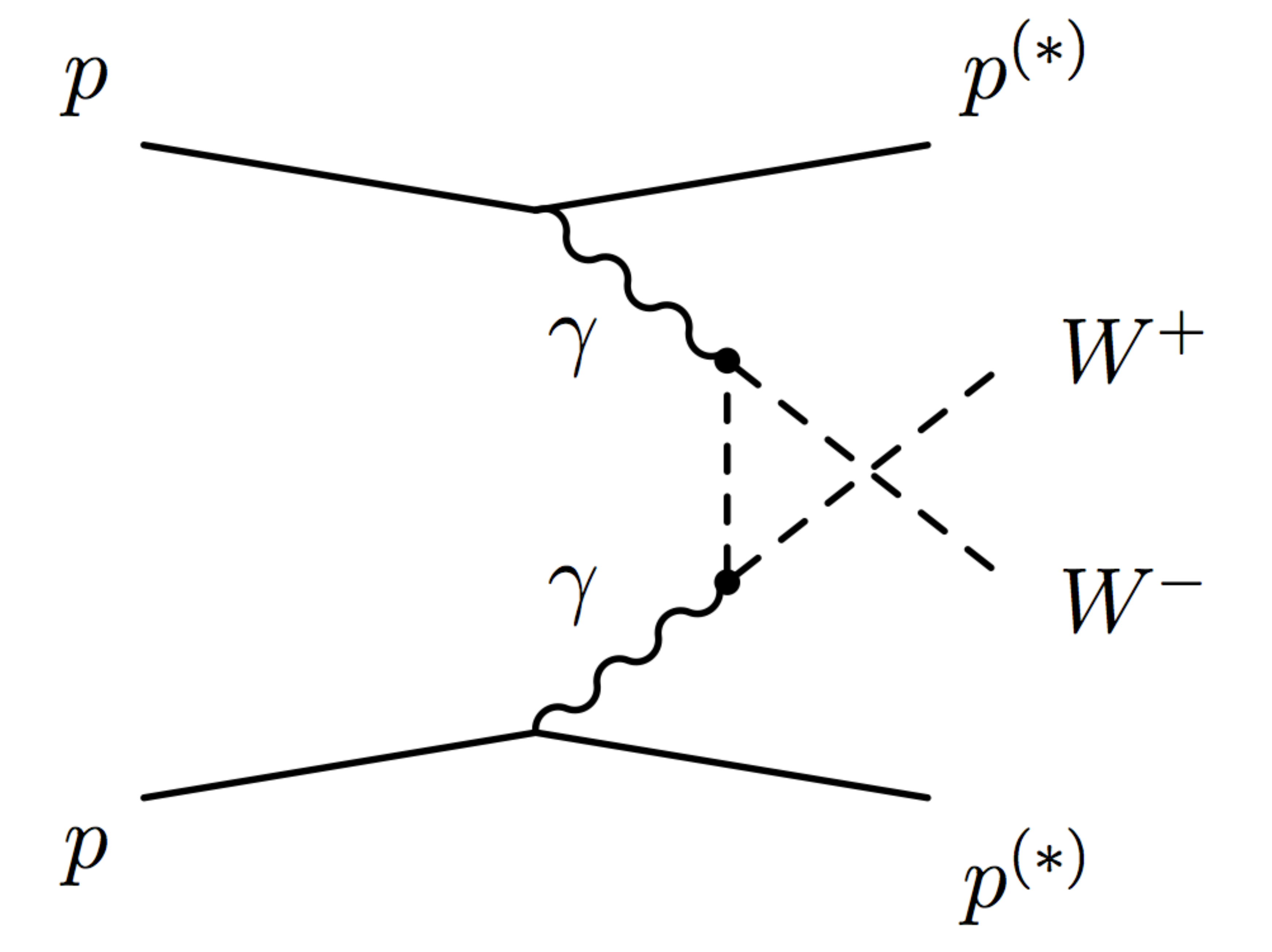}\hfill
\caption{Quartic (left), $t$-channel (center), and $u$-channel (right) diagrams contributing to the $\PGg\PGg \to \PWp\PWm$ process at leading order in the SM. The $\Pp^{(*)}$ indicates that the final state proton(s) remain intact (``exclusive'' or ``elastic'' production), or dissociate (``quasi-exclusive'' production).
\label{fig:FeynmanDiagrams}}
\end{figure}

A nonnegligible fraction of proton-proton collisions at the CERN LHC involves (quasi-real) photon interactions that provide a unique opportunity
to study high-energy $\PGg\PGg$ processes at center-of-mass energies and integrated luminosities much higher than previously available~\cite{deFavereaudeJeneret:2009db}.
Using the $\sqrt{s} = 7$\TeV data collected during Run 1 of the LHC, where Run 1 refers to the LHC data collection period between 2010-2012, measurements of $\PGg\PGg \to \PGmp\PGmm$~\cite{Chatrchyan:2011ci,Aad:2015bwa} and $\PGg\PGg \to \Pep\Pem$~\cite{Chatrchyan:2012tv,Aad:2015bwa} production were performed, followed by the first stu\-dies of $\PGg\PGg \to \PWp\PWm$~\cite{Chatrchyan:2013akv}. The latter process, occurring at leading order via the diagrams shown in Fig.~\ref{fig:FeynmanDiagrams}, is particularly well suited to search for physics beyond the standard model (SM). Such deviations from the SM may be quantified through anomalous quartic gauge couplings (AQGC) of operators of dimension-6 or -8~\cite{Pierzchala:2008xc,Chapon:2009hh}. Specific models including anomalous gauge-Higgs couplings~\cite{Maniatis:2008zz,Delgado:2014jda}, as well as composite Higgs~\cite{Delgado:2014jda,Fichet:2013ola,Espriu:2014jya} or warped extra dimensions~\cite{Fichet:2013ola} scenarios, will also result in deviations from the SM predictions for the $\PGg\PGg \to \PWp\PWm$ (differential and/or integrated) cross sections. Prior to the LHC, limits on AQGC were obtained through triboson ($\Z\PGg\PGg$ and $\PWp\PWm\PGg$) production, and $\PW\PW \to \PGg\PGg$ scattering at LEP~\cite{Heister:2004yd,Abbiendi:2004bf,Abbiendi:2003jh,Abbiendi:1999aa,Abdallah:2003xn,Achard:2002iz,Achard:2001eg}, and through $\PGg\PGg \to \PWp\PWm$ scattering at the Tevatron~\cite{Abazov:2013opa}. Anomalous quartic gauge couplings have been explored at the LHC through triboson ($\PW\PGg\PGg$ or $\PW V\PGg$, where V is a $\PW$ or $\Z$ boson) production~\cite{Chatrchyan:2014bza,Aad:2015uqa}, and same-charge $\PW\PW \to \PW\PW$ scattering~\cite{Aad:2014zda,Khachatryan:2014sta}.

This paper presents an update of the 7\TeV CMS $\PGg\PGg \to \PWp\PWm$ measurement~\cite{Chatrchyan:2013akv}, largely following the same analysis strategy as for 7\TeV but using the 8\TeV data set collected in 2012. The signal topology considered is $\Pp\Pp \to \Pp^{(*)}\PWp\PWm\Pp^{(*)}$, where the $\Pp^{(*)}$ indicates that the final state protons either remain intact (``exclusive'' or ``elastic'' production), or dissociate into an undetected system (``quasi-exclusive'' or ``proton dissociation'' production). The  $\PWp\PWm \to \PGm^{\pm}\Pe^{\mp}$  (plus undetected neutrinos) channel is the final state used to search for a signal, as the backgrounds due to Drell--Yan (DY) and  $\Pgg\Pgg \to\Plp\Plm$ production are smaller than in the same-flavor final states. Events in which one or both of the $\PW$ bosons decay into a tau lepton, with a subsequent decay of the tau to a muon or electron and neutrinos, are also included in the signal. In contrast to exclusive production, inclusive $\PWp\PWm$ production is always accompanied by underlying event activity originating from semihard multiple-parton interactions and from softer ``spectator" partons at forward rapidities. This will almost always result in the production of additional detectable charged particles from the $\PGm^{\pm}\Pe^{\mp}$ vertex.
The experimental signature for the signal therefore consists of a muon-electron pair with large transverse momentum $\pt(\PGm^{\pm}\Pe^{\mp})$, the vector $\pt$ sum of the pair, originating from a common primary vertex with no additional charged particles detected.

Control samples of $\PGg\PGg \to\PGmp\PGmm$ and $\PGg\PGg \to \Pep\Pem$ events are used to study the efficiency of the exclusive selection in data, as well as the ``rescattering" corrections~\cite{Khoze:2000db,HarlandLang:2012qz}, from additional parton interactions between the protons, in quasi-exclusive collisions.
Control regions in the dilepton $\pt$ and charged-particle multiplicity distributions are used to study the main background contributions to the signal. Finally the $\pt(\PGm^{\pm}\Pe^{\mp})$ distribution is used as the discriminating variable to measure the standard model $\PGg\PGg \to \PWp\PWm$ cross section, and to search for evidence of AQGC.

Sections 2--3 give a general description of the theory and the CMS detector, while Sections 4--5 describe the data sets, Monte Carlo (MC) simulations, and event selection. Sections 6--8 explain the 8\TeV analysis, and Section 9 describes the present 8\TeV results, as well as their combination with those from the previous 7\TeV study.

\section{Phenomenology of anomalous couplings in \texorpdfstring{$\PGg\PGg \to \PWp\PWm$}{gamma-gamma to W+W-} interactions}
\label{tab:AQGCTheory}

Within the SM, the triple ($\PW\PW\PGg$) and quartic ($\PW\PW\PGg\PGg$) couplings that contribute to $\PGg\PGg \to \PWp\PWm$ production are fully connected through the requirement of gauge invariance. In contrast, effective field theories can be constructed to quantify potential deviations from the SM by introducing genuine AQGCs through dimension-6 operators that are not related to the SM triple or quartic couplings~\cite{Belanger:1992qh}. By imposing $U(1)_{\mathrm{EM}}$ and global custodial $SU(2)_{C}$ symmetries, and further requiring charge-conjugation and parity to be separately conserved, two such operators are allowed with couplings denoted $a_{0}^{\PW}/\Lambda^{2}$ and $a_{C}^{\PW}/\Lambda^{2}$, where $\Lambda$ is the energy scale for new physics. This approach corresponds to assuming a nonlinear representation of the spontaneously broken $SU(2) \otimes U(1)$ symmetry.

With the discovery of a light Higgs boson~\cite{Aad:2012tfa,Chatrchyan:2012ufa,Chatrchyan:2013lba}, a linear realization of the $SU(2) \otimes U(1)$ symmetry of the SM, spontaneously broken by the Higgs mechanism, is possible. Thus, the lowest order operators, where new physics may cause deviations in the quartic gauge boson couplings alone, are of dimension 8. In the dimension-8 formalism~\cite{Belanger:1999aw,Eboli:2006wa,Baak:2013fwa} there are fourteen operators contributing to $\PW\PW\PGg\PGg$ couplings, which in general will also generate a $\PW\PW\Z\Pgg$ vertex. By assuming that the anomalous $\PW\PW\Z\PGg$ vertex vanishes, a direct relationship between the dimension-8 $f_{{M,0,1,2,3}}/\Lambda^{4}$ couplings and the dimension-6 $a_{{0,C}}^{\PW}/\Lambda^{2}$ couplings can be recovered~\cite{Belanger:1999aw,Baak:2013fwa,Eboli:2006wa,Chatrchyan:2014bza}:
\begin{equation}
\begin{aligned}
\dfrac{a_{0}^{\PW}}{\Lambda^{2}}&=-\dfrac{4M_{\PW}^{2}}{e^{2}}\dfrac{f_{{M,0}}}{\Lambda^{4}},\\
\dfrac{a_{C}^{\PW}}{\Lambda^{2}}&=\dfrac{4M_{\PW}^{2}}{e^{2}}\dfrac{f_{{M,1}}}{\Lambda^{4}},\\
\end{aligned}
\end{equation}
where $M_\PW$ is the mass of the \PW~boson and $e$ is the unit of electric charge. The $f_{{M,2,3}}/\Lambda^{4}$ couplings
can be determined from the relations
$f_{{M,0}} = 2  f_{{M,2}}$
and $f_{{M,1}} = 2  f_{{M,3}}$,
which are a result of the constraint on the $\PW\PW\Z\PGg$ vertex vanishing.

In both dimension-6 and dimension-8 scenarios, the $\PGg\PGg \to \PWp\PWm$ cross section in the presence of anomalous couplings would increase rapidly with the photon-photon center-of-mass energy $W_{\PGg\PGg}$. For couplings of the size that can be probed with the current data set, this would result in violation of unitarity at scales well below those reached in 7 and 8\TeV $\Pp\Pp$ collisions at the LHC. To prevent this, various approaches modifying the effective Lagrangian have been proposed~\cite{Eboli:2000ad,Gupta:2011be,Alboteanu:2008my}. In this analysis, following the previous $\PGg\PGg \to \PWp\PWm$ results from the LHC and Tevatron, we consider a dipole form factor with a cutoff scale $\Lambda_{\text{cutoff}}$:
\begin{equation*}
a^{\PW}_{0,C}(W^{2}_{\PGg\PGg}) = \frac{a^{\PW}_{0,C}}{\left( 1+\frac{W^{2}_{\PGg\PGg}}{\Lambda_{\text{cutoff}}^{2}} \right)^{2}}.
\end{equation*}

We quote both the limits which preserve unitarity, with a dipole form factor and $\Lambda_{\text{cutoff}} = 500$\GeV as was used in previous publications~\cite{Chatrchyan:2013akv,Abazov:2013opa}, and limits with $\Lambda_{\text{cutoff}} \to \infty$, which is equi\-valent to no form factor, violating unitarity.

The presence of anomalous couplings among the gauge bosons is expected to result in a harder spectrum of the transverse momentum of the \PW-pair system which can be probed experimentally by the hardness of the spectra in their decay products and, suitably, by that of the dilepton system of electron and muon.

\section{The CMS detector}

A detailed description of the CMS experiment can be found in Ref.~\cite{JINST}. The central feature of the CMS apparatus is a superconducting solenoid, of 6~m internal diameter.  Within the field volume are the silicon pixel and strip tracker, the crystal electromagnetic calorimeter (ECAL) and the brass-scintillator hadronic calorimeter (HCAL).  Muons are measured in gas-ionization detectors embedded in the steel flux-return yoke of the solenoid.
The momentum resolution for electrons with
$\pt \sim  45$\GeV
from $\Z \to \Pe \Pe$ decays ranges from 1.7\% for nonshowering
electrons in the barrel region to 4.5\% for showering electrons in the
endcaps~\cite{Khachatryan:2015hwa}.
The calorimeter cells are grouped in projective towers, of granularity $\Delta\eta{\times}\Delta\phi=0.087{\times}0.087$ (where $\phi$ is the azimuthal angle in radians) in the pseudorapidity region $\abs{\eta}<1.5$, and increasing
to 0.175$\times$0.175 in the region $3<\abs{\eta}<5$. The silicon tracker covers a range of $\abs{\eta}<2.4$, and
consists of three layers made of 66~million $100{\times}150\mum^{2}$ pixels followed by ten microstrip layers, with strips of pitch between 80 and $180\mum$. The silicon tracker is used to detect charged particles as tracks.
Muons are measured in the window $\abs{\eta}<2.4$, with detection planes made of three technologies: drift
tubes, cathode strip chambers, and resistive-plate chambers. Thanks to the strong magnetic field, 3.8\unit{T}, and to the high granularity of the silicon
tracker, the transverse momentum, \pt, of the muons matched to silicon tracks is measured with a resolution better than 1.5\%, for \pt smaller than 100\GeV.
The resolution of $z_0$, the point of closest approach of the track to the beam direction $z$, for a 1 (10)\GeV pion is 100--300 $\mu$m (30--60 $\mu$m) in the central region and 300--1000 $\mu$m (60--150 $\mu$m) in the forward region~\cite{Chatrchyan:2014fea}.
The ECAL provides coverage in a range of $\abs{ \eta }< 1.48 $ in a barrel region, and $1.48 <\abs{ \eta } < 3$ in two endcap regions (EE).
A preshower detector consisting of two planes of silicon sensors interleaved with a total of 3 radiation lengths of lead is located in front of the EE.
The first level of the CMS trigger system, composed of custom hardware processors, uses information from the calorimeters and muon detectors to
select (in less than 4\mus) the most interesting events. The high-level trigger processor farm further decreases the event
rate from 100\unit{kHz} to a few hundred Hz, before data storage.

\section{Data sets and Monte Carlo simulation}
\label{sec:DatasetsMC}

The analyzed data samples consist of 19.7\fbinv of proton-proton collisions collected in 2012 at a center-of-mass energy of $\sqrt{s} = 8$\TeV. This measurement is combined with a similar analysis carried out in 2011 using 5.05\fbinv of $\Pp\Pp$ collisions collected at a center-of-mass energy of $\sqrt{s} = 7$\TeV. During Run 1 of the LHC the number of overlapping interactions per bunch crossing (``pileup'')
was nonnegligible. In the 8 (7)\TeV data-taking period the average pileup was 21 (9) interactions per crossing. The 7\TeV data analysis is described in Ref.~\cite{Chatrchyan:2013akv}, and the rest of this
section focuses on describing the 8\TeV data analysis.

The simulated SM and anomalous signal samples of $\PGg\PGg \to \PWp\PWm$ are generated with \MADGRAPH~\cite{Alwall:2011uj,Alwall:2014hca} v5 release 2.0.0, using the equivalent photon approximation (EPA)~\cite{Budnev:1974de}. Cross-checks with \CALCHEP~\cite{Belyaev:2012qa} v3.4 have also been performed since this is the generator used for simulated SM and anomalous signal samples for the 7\TeV analysis. The elastic and proton dissociation $\PGg\PGg \to\Plp\Plm$ samples are produced using the \textsc{lpair} v4.0 generator~\cite{Vermaseren:1982cz,Baranov:1991yq}.

The backgrounds from inclusive diboson, W+jets, and \ttbar production are simulated with \MADGRAPH.
For \ttbar production the yields are normalized to the next-to-next-to-leading-order (NNLO) plus next-to-next-to-leading-logarithmic cross section prediction obtained with Top++2.0~\cite{Czakon:2011xx}.
For inclusive diboson and W+jets production the yields are normalized to the NNLO and next-to-leading-order cross section predictions, respectively,
and are obtained with \MCFM v6.6~\cite{Campbell:2010ff}.
Inclusive Drell--Yan samples are simulated with \POWHEG v1.0~\cite{Nason:2004rx,Frixione:2007vw,Alioli:2010xd}.
The outgoing partons from the matrix element calculation in both  \MADGRAPH and \POWHEG are matched to parton showers from the \PYTHIA v6.4.26~\cite{Sjostrand:2006za} with the Z2* tune~\cite{Chatrchyan:2013gfi} for
the analysis of the 8\TeV data and with the Z2 tune~\cite{Field2011} for the analysis of the 7\TeV data.
The simulated inclusive $\PWp\PWm$ sample does not include events generated in diffractive topologies, in which one of the incoming protons remains intact.
While diffractive $\PWp\PWm$ production is expected to be small compared to the rate of inclusive $\PWp\PWm$ production, the mean multiplicity of charged particles in diffractive events will be smaller, thus enhancing the contribution of diffractive production to the exclusive signal region. The contribution from diffractive $\PWp\PWm$ production is simulated with \textsc{pompyt} v2.6.1~\cite{Bruni:1993is}, using diffractive parton distribution functions obtained from the H1 fit B to diffractive deep inelastic scattering data~\cite{Aktas:2006hy}. In practice, the diffractive $\PWp\PWm$ background may be suppressed by a ``gap survival probability'' factor, representing the effect of rescattering interactions that will lead to additional hadronic activity in the event. As this factor is not precisely predicted or measured at LHC energies~\cite{Chatrchyan:2012vc}, a very conservative 100\% gap survival probability (meaning no correction due to rescattering interactions) is assumed.
Gluon-induced central exclusive production of $\PWp\PWm$ pairs, with an additional ``screening'' gluon emission to cancel the color flow, is expected to be heavily suppressed~\cite{Lebiedowicz:2012gg} and is neglected in the current analysis.

Electroweak production (at order $\alpha_\mathrm{EW}^{5}$ or $\alpha_\mathrm{EW}^{6}$ for real $\PWp\PWm$ emission) of $\PWp\PWm$ pairs, including $\PW\PW\to \PW\PW$ scattering, is also not included in the simulated inclusive $\PWp\PWm$ sample. We use a sample generated with \MADGRAPH, which describes well the electroweak production of $\Z\PQq\PQq$ (where `q' indicates a quark jet) at $\sqrt{s}=8$\TeV~\cite{Khachatryan:2014dea}, to estimate the central value of the electroweak $\PWp\PWm\PQq\PQq$ background, with \PHANTOM v1.0~\cite{Ballestrero:2007xq} used for systematic studies.

All simulated samples are passed through a detailed \GEANTfour simulation~\cite{Agostinelli:2002hh} of the CMS detector.
The same algorithms are used to reconstruct both the simulated samples and collision data.

\section{Event selection}
\label{sec:EventSelection}

The event selection is similar for both $\PGm^{\pm}\Pe^{\mp}$ final states used to search for a $\PWp\PWm$ signal, and for the
$\PGmp\PGmm$ and $\Pep\Pem$ final states used as control samples. The events are triggered by the presence of two leptons with transverse momentum $\pt > 17 (8)$\GeV for the leading (subleading) lepton.

Offline, the leptons are required to be of opposite charge, to have $\pt > 20$\GeV,
to pass ``tight'' identification criteria for muons~\cite{Chatrchyan:2012xi}
and ``medium'' identification criteria for electrons~\cite{Khachatryan:2015hwa}, and
come from the same reconstructed primary vertex.
Primary vertices are identified by clustering tracks according to a deterministic annealing algorithm, and
subsequently performing a fit to the clustered tracks~\cite{Chatrchyan:2014fea}.
The ``tight'' muon identification includes requirements on the minimum number of muon detector planes hit, the minimum number of hits in the pixel detector and of layers hit in the silicon strip detector, the goodness of the global fit to the muon track, and the transverse impact parameter with respect to the primary vertex.
An additional requirement that the longitudinal impact parameter be at most 5~mm is added for the 8\TeV analysis and was not present in the 7\TeV analysis. This requirement is added to suppress cosmic muons, muons from decays in flight of charged mesons, and tracks
from pileup.
The ``medium'' identification criteria for electrons combines
information from the ECAL, HCAL, and silicon tracker.
This includes selections on the azimuthal angle and pseudorapidity differences between the tracks and ECAL deposits associated to the electron
candidate, the ratio of energy deposited in the HCAL to that in the ECAL, the shower shape of the ECAL deposits, and the compatibility of the energy
deposited in the ECAL with the momentum of the associated track. The efficiency of the ``tight'' muon identification is estimated to be $\geq$96\% for muons with $\pt > 20$\GeV, with a hadron misidentification probability of $<$0.5\%. The efficiency of the ``medium'' electron identification rises from $\sim$60\% for electrons with $\pt = 20$\GeV, reaching a plateau at $\geq$80\% for electrons with $\pt > 55$\GeV. The misidentification probability is estimated to be $<$4\% for the ``medium'' electron identification.
In addition, the electrons are required to satisfy relative isolation
criteria, based on the global particle-flow algorithm~\cite{CMS:2009nxa,CMS:2010nxa}. The invariant dilepton mass is also required to satisfy $m(\Plp\Plm) > 20$\GeV in order to remove any potential background due to low-mass resonances, which is particularly relevant in the $\PGmp\PGmm$ and $\Pep\Pem$ final states.

The final signal region is then defined by the presence of an opposite-charge electron-muon pair originating from a common primary vertex that has no additional tracks associated with it, and has transverse momentum $\pt(\PGm^{\pm}\Pe^{\mp}) > 30$\GeV. The zero-additional-tracks requirement is motivated by the lack of
underlying event activity expected for exclusive and quasi-exclusive $\PGg\PGg \to \PWp\PWm$ production, in which the beam protons remain intact or dissociate into an
undetected forward system respectively, in contrast to backgrounds from inclusive diboson production. The $\pt(\PGm^{\pm}\Pe^{\mp}) > 30$\GeV requirement is designed to suppress
backgrounds from $\PGt^{+}\PGt^{-}$ production, including the exclusive and quasi-exclusive $\PGg\PGg \to \PGt^{+}\PGt^{-}$ processes.

\section{The \texorpdfstring{$\PGg\PGg\to\Plp\Plm$}{gamma-gamma to l+l-} control samples and corrections}
\label{sec:ControlSamplesCorrections}

In the $\PGmp\PGmm$ and $\Pep\Pem$ final states, backgrounds due to direct $\PGg\PGg \to\Plp\Plm$
production and Drell--Yan processes are much larger than in the $\PGm^{\pm}\Pe^{\mp}$ channel. Therefore these channels are used as control samples to study both the efficiency of the zero-additional-tracks selection, and the theoretically poorly
known proton dissociation contribution to high-mass $\PGg\PGg$ interactions~\cite{daSilveira:2014jla}.

\subsection{Efficiency correction for track veto}
\begin{figure}[htb]
\centering
\includegraphics[width=.40\textwidth]{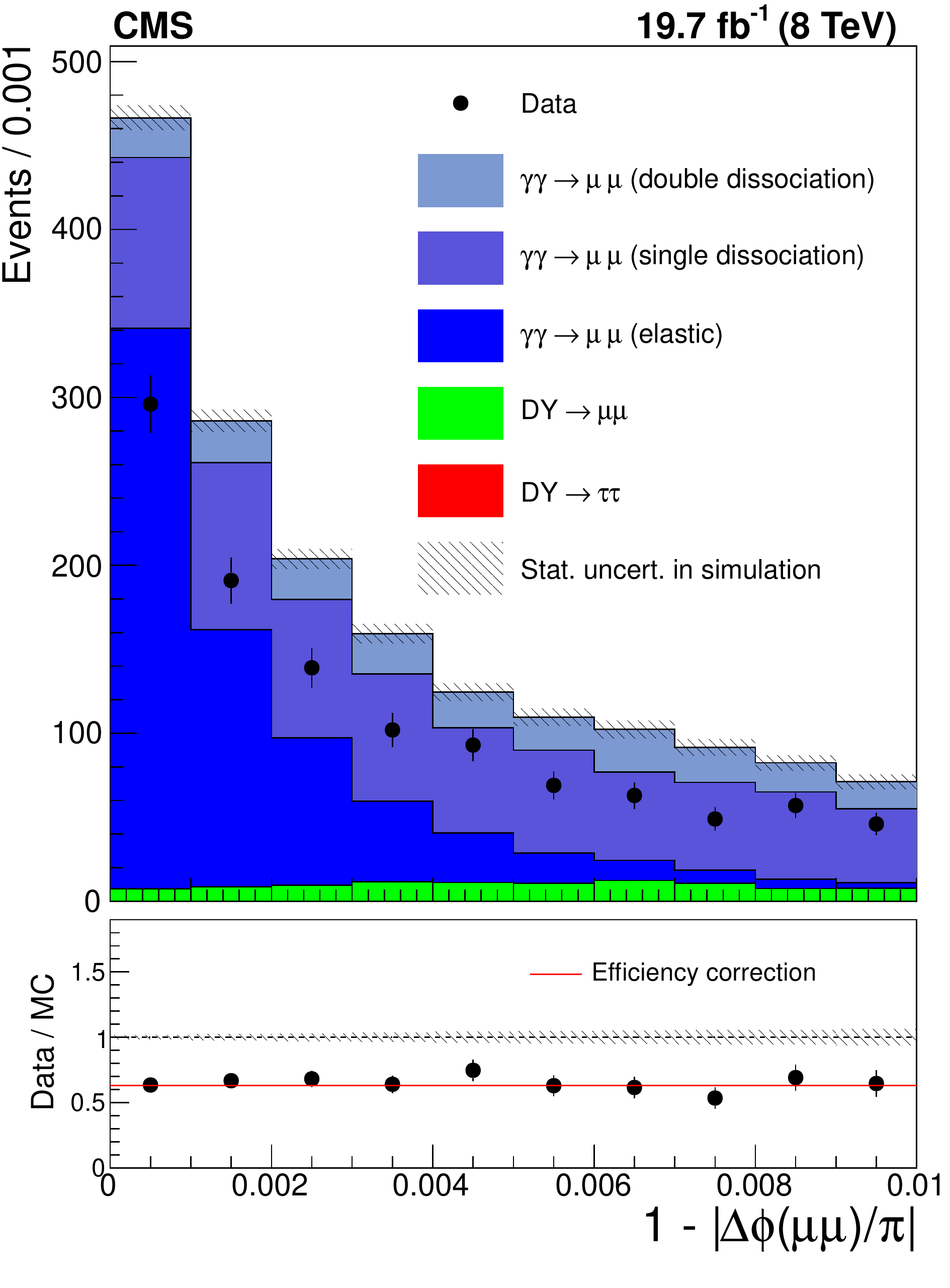}
\includegraphics[width=.40\textwidth]{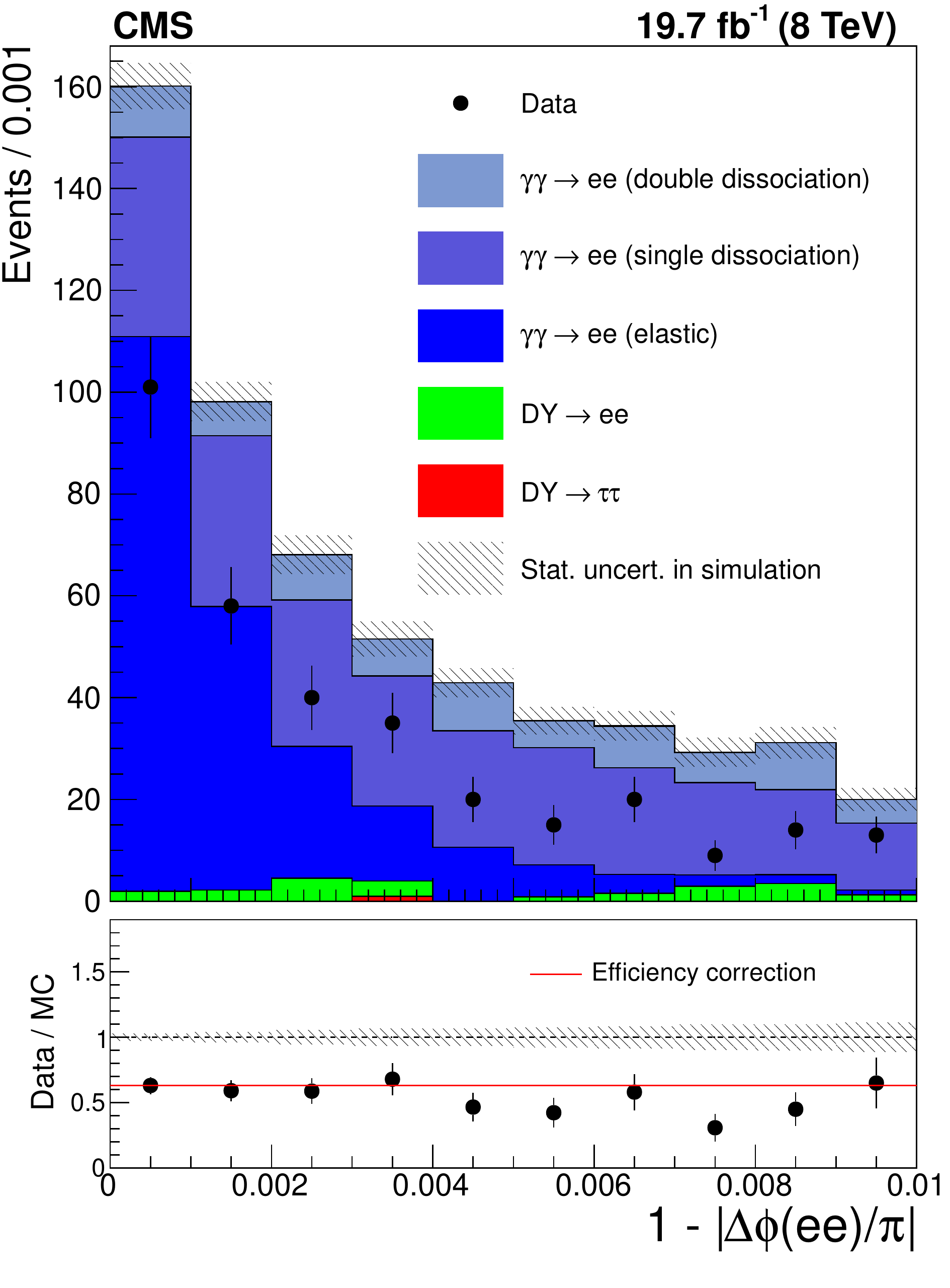}
\caption{Acoplanarity for the $\PGmp \PGmm$ (left) and $\Pep\Pem$ (right) final states in the elastic $\PGg\PGg\to\Plp\Plm$ control region ($\abs{1-\Delta\phi(\Plp\Plm)/\pi}<0.01$ and $m(\Plp\Plm) < 70$\GeV or $m(\Plp\Plm) > 106$\GeV) and 0 additional tracks associated to the dilepton vertex.
The data (points with error bars) are compared to the simulated samples (histograms) in the top panels. The data/MC ratios are shown in the bottom panels (the red line shows the extracted correction for the track veto efficiency).
\label{fig:elasticggllplots}}
\end{figure}

\begin{figure}[htb]
\centering
\includegraphics[width=.40\textwidth]{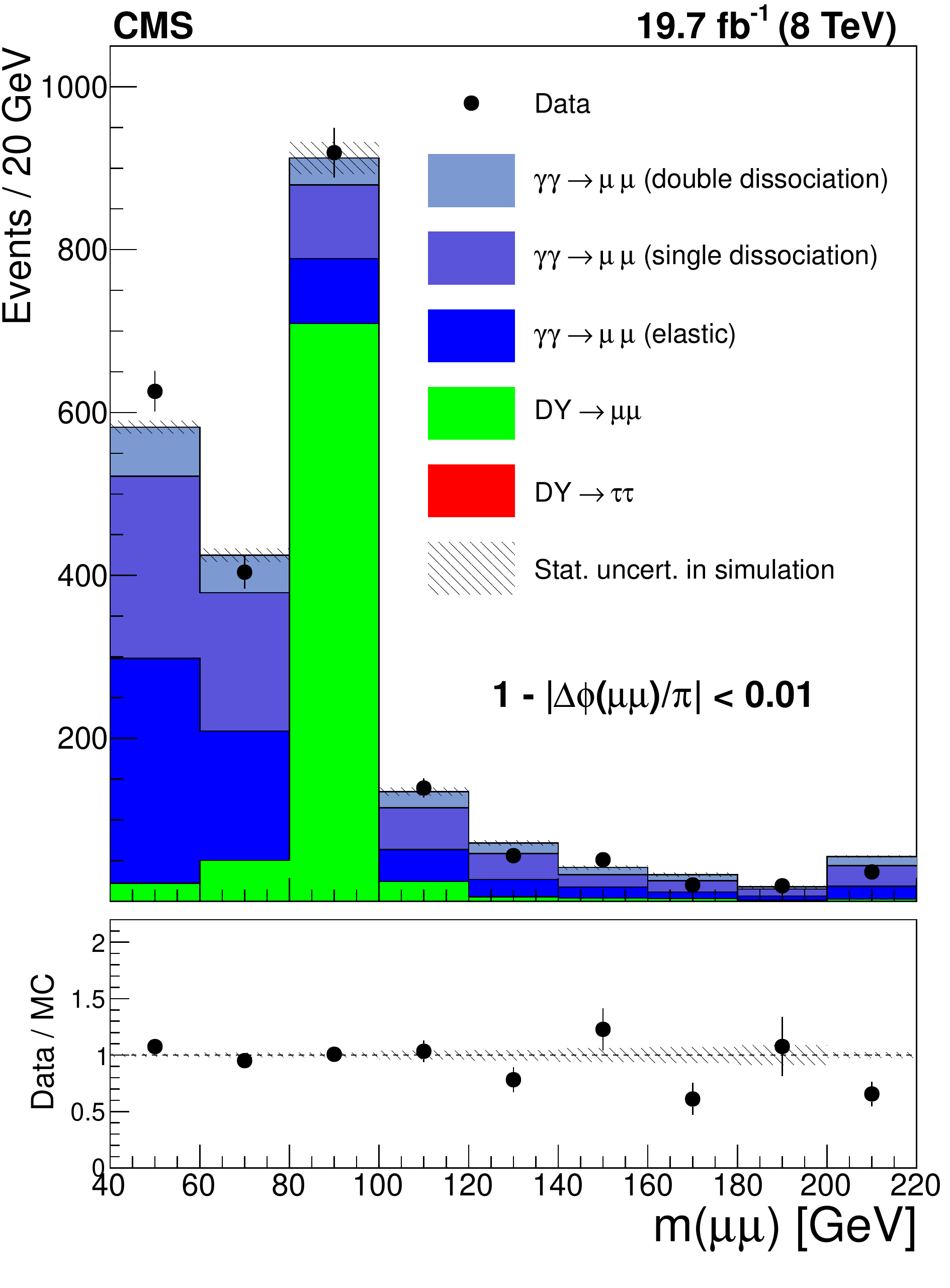}
\includegraphics[width=.40\textwidth]{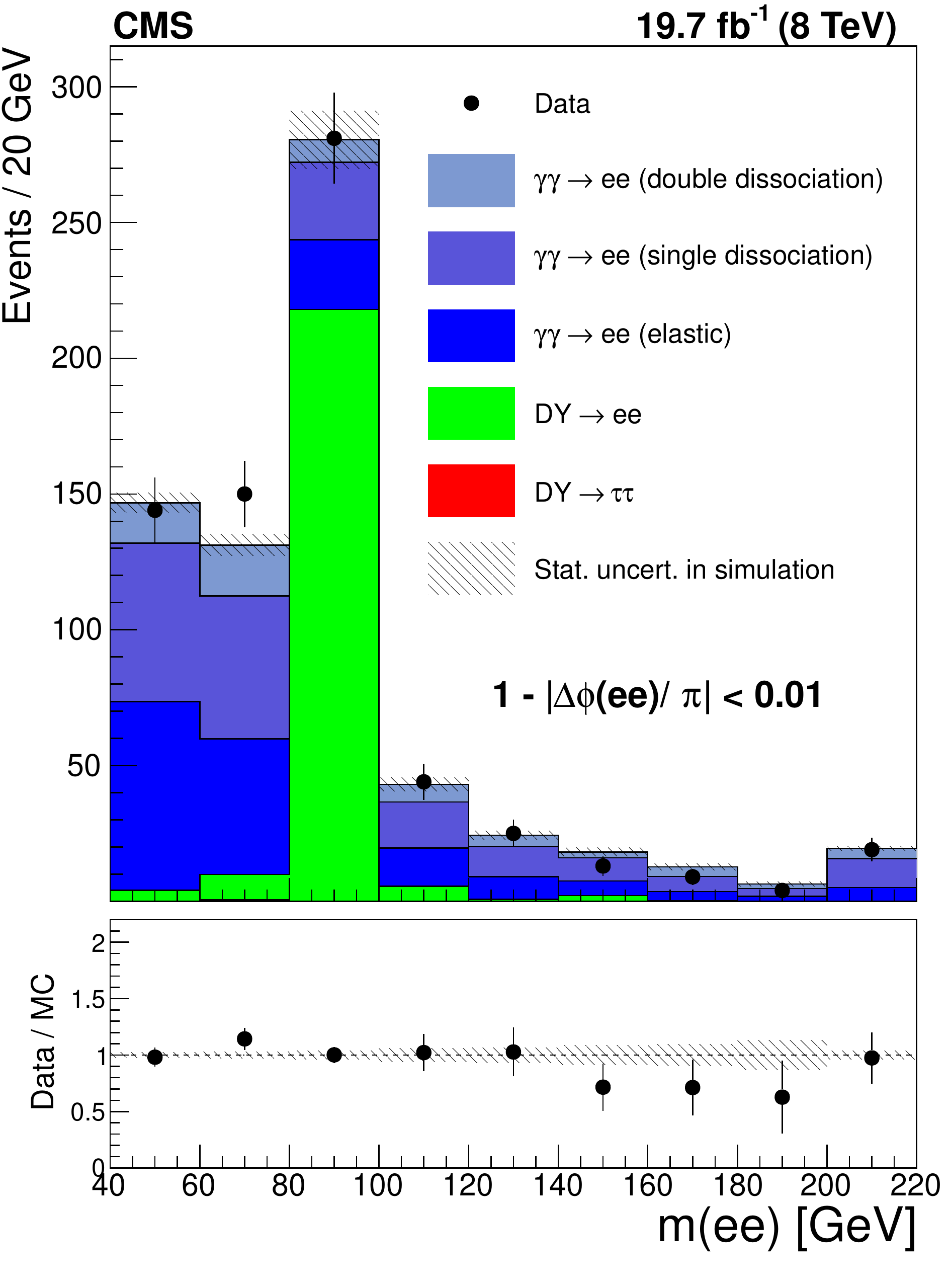}
\caption{Dilepton invariant mass for the $\PGmp \PGmm$ (left) and $\Pep \Pem$ (right) final states
with an acoplanarity requirement, $\abs{1-\Delta\phi(\Plp\Plm)/\pi}<0.01$,
and zero additional tracks associated to the dilepton vertex.
The data (points with error bars) are compared to the simulated samples (histograms) in the top panels, and the data/MC ratios are shown in the bottom panels. The exclusive-production simulated samples are scaled to the number of events in data for $m(\Plp\Plm) < 70$\GeV or $m(\Plp\Plm) > 106$\GeV. The Drell--Yan simulation is scaled to the number of events in data for $70 < m(\Plp\Plm) < 106$\GeV. The last bin in both plots is an overflow bin and includes all events with invariant mass greater than 200\GeV.
\label{fig:elasticggllMassplots}}
\end{figure}

First, in order to select a high-purity sample of elastic $\Pp\Pp \to \Pp \Plp\Plm \Pp$ events and study the efficiency of the
additional track veto, we apply
harsh selection criteria to the kinematics of the lepton pair. These consist in requiring a small acoplanarity, $\abs{1-\Delta\phi(\Plp\Plm)/\pi}<0.01$ where $\Delta\phi(\Plp\Plm)$ is the difference in azimuthal angle between the two leptons, and an invariant mass incompatible with $\Z \to \Plp\Plm$ decays ($m(\Plp\Plm) < 70$\GeV or $m(\Plp\Plm) > 106$\GeV).
The leptons from elastic $\Pp\Pp \to \Pp \Plp\Plm \Pp$ events have small acoplanarity because the very small virtuality of the exchanged photons results in a dilepton pair produced with $\pt(\Plp\Plm)\sim 0$.
The acoplanarity and invariant mass requirements result in a sample expected to contain a negligible contribution from inclusive backgrounds, but some contamination from $\PGg\PGg \to \Plp\Plm$ events where one or both of the protons dissociate.
In this control sample we find a deficit in the data compared to the theoretical prediction for events
with zero additional tracks associated to the dilepton vertex (Fig.~\ref{fig:elasticggllplots}). We have verified that this is due to the fact that the efficiency of the additional-track veto is overestimated in the simulation. To numerically calculate the data-to-simulation ratio, we use a tighter acoplanarity requirement ($\abs{1-\Delta\phi(\Plp\Plm)/\pi}<0.001$, corresponding to ${>}3\sigma$ in terms of the experimental resolution
on the acoplanarity) to further reduce the contamination from $\PGg\PGg \to\Plp\Plm$ events where one or both of the protons dissociate.
The data-to-simulation ratio is $0.63 \pm 0.04$ in the $\PGmp\PGmm$ channel and $0.63 \pm 0.07$ in the $\Pep\Pem$ channel.
By comparing the shapes of the $\PGg\PGg\to\Plp\Plm$ distributions we find a good data-to-theory agreement, apart from the overall difference in normalization (Figs.~\ref{fig:elasticggllplots} and~\ref{fig:elasticggllMassplots}). We therefore apply this ratio, averaged over the $\PGmp\PGmm$
and $\Pep\Pem$ samples, as a track veto efficiency correction to the $\PGg\PGg \to \PWp\PWm$ signal.

In exclusive and quasi-exclusive production, additional tracks identified as coming from the dilepton vertex are predominantly misassociated tracks originating from other pileup vertices in the event. These are mainly very low-$\pt$, forward tracks not modeled
perfectly by the simulation. Therefore a track veto efficiency correction is applied to account for the resulting migration of signal events to higher multiplicities.
For inclusive backgrounds, migrations may happen in both directions, with tracks from pileup vertices being wrongly associated with a dilepton vertex, or
tracks from the underlying event being wrongly associated with a pileup vertex. For the largest background of inclusive $\PWp\PWm$ production, the simulation is observed to
reproduce the data in the relevant control region; therefore no correction is applied to the backgrounds.

\subsection{Proton dissociation contribution}
Simulations of high-mass $\PGg\PGg$ interactions with the \textsc{lpair} matrix element generator show that they predominantly occur in events where at least one of the protons dissociates. However, the cross section calculations do not include rescattering effects, in which additional
gluon interactions between the protons produce extra hadronic activity in the event besides the final-state leptons or gauge bosons. As a result of
these rescattering corrections, $\PGg\PGg\to\Plp\Plm$ and $\PGg\PGg\to \PWp\PWm$ signal events will migrate to
higher multiplicities.
This is expected to be a large effect, particularly for events in which both protons dissociate, with up to about 90\% of
events being nonexclusive, depending on the exact kinematic range studied~\cite{HarlandLang:2012qz,daSilveira:2014jla}.
The contribution from proton dissociation is therefore estimated directly from the data, rather than relying on simulation.

To estimate the contribution due to proton dissociation in a kinematic region similar to the $\PWp\PWm$ signal, we select a sample of dilepton events with invariant mass greater than 160\GeV, corresponding to the threshold for the production of two on-shell W bosons, with no additional tracks associated with the dilepton vertex.
We then compute the ratio of the number of events measured in this region
to the predicted number of elastic $\Pp\Pp \to \Pp \Plp\Plm \Pp$ events, with the additional track veto efficiency correction applied and the
Drell--Yan contribution subtracted from the data. This results in a scale factor $\mathrm{F} = 4.10\pm 0.43$, with the uncertainty determined from the statistical uncertainty of the data control sample, that is used to correct the elastic $\Pp\Pp \to \Pp \PWp\PWm \Pp$ prediction to the total $\Pp\Pp \to \Pp^{(*)} \PWp\PWm \Pp^{(*)}$ prediction, including proton dissociation.

\begin{figure}[htb]
\centering
\includegraphics[width=.45\textwidth]{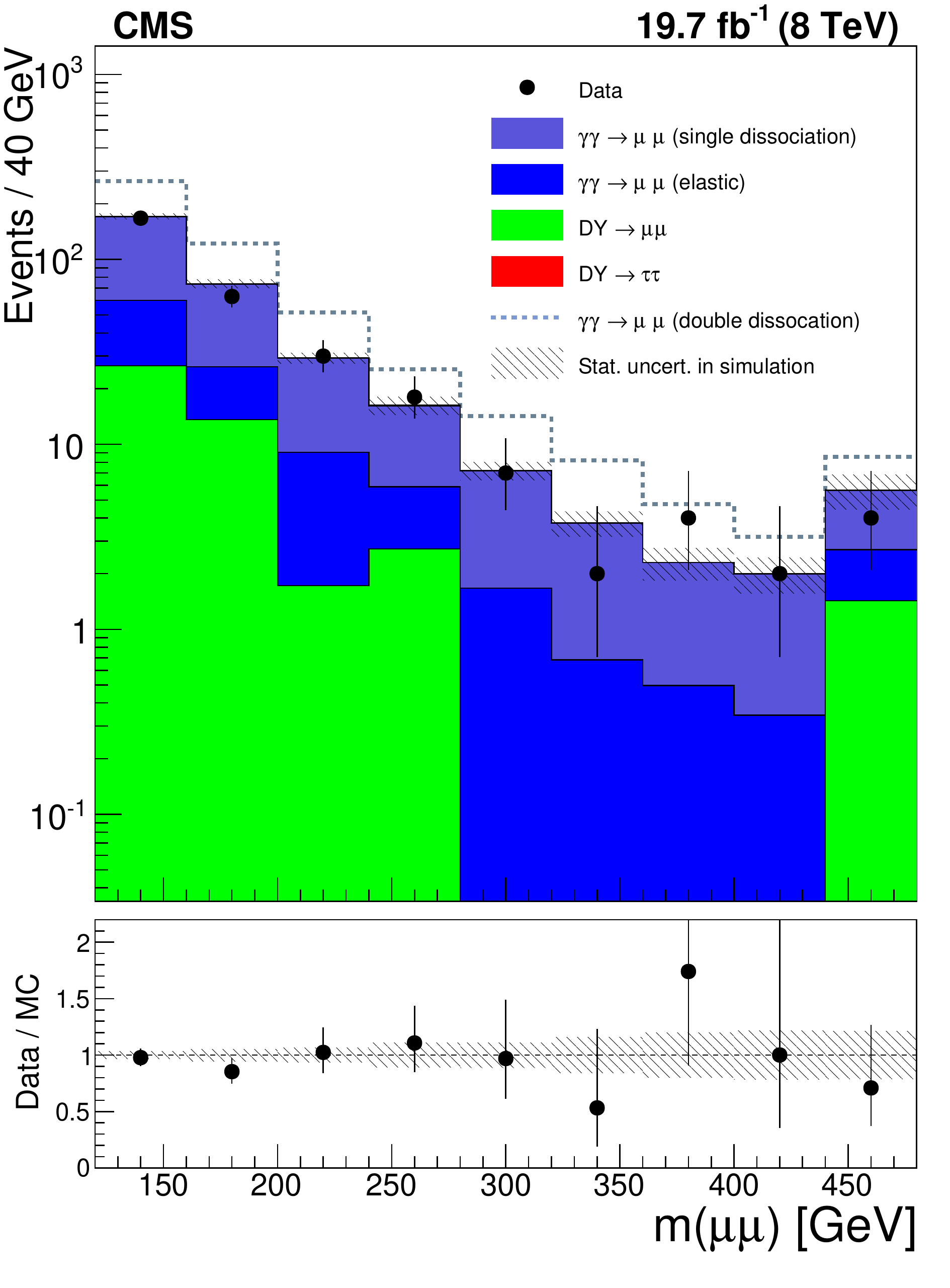}
\includegraphics[width=.45\textwidth]{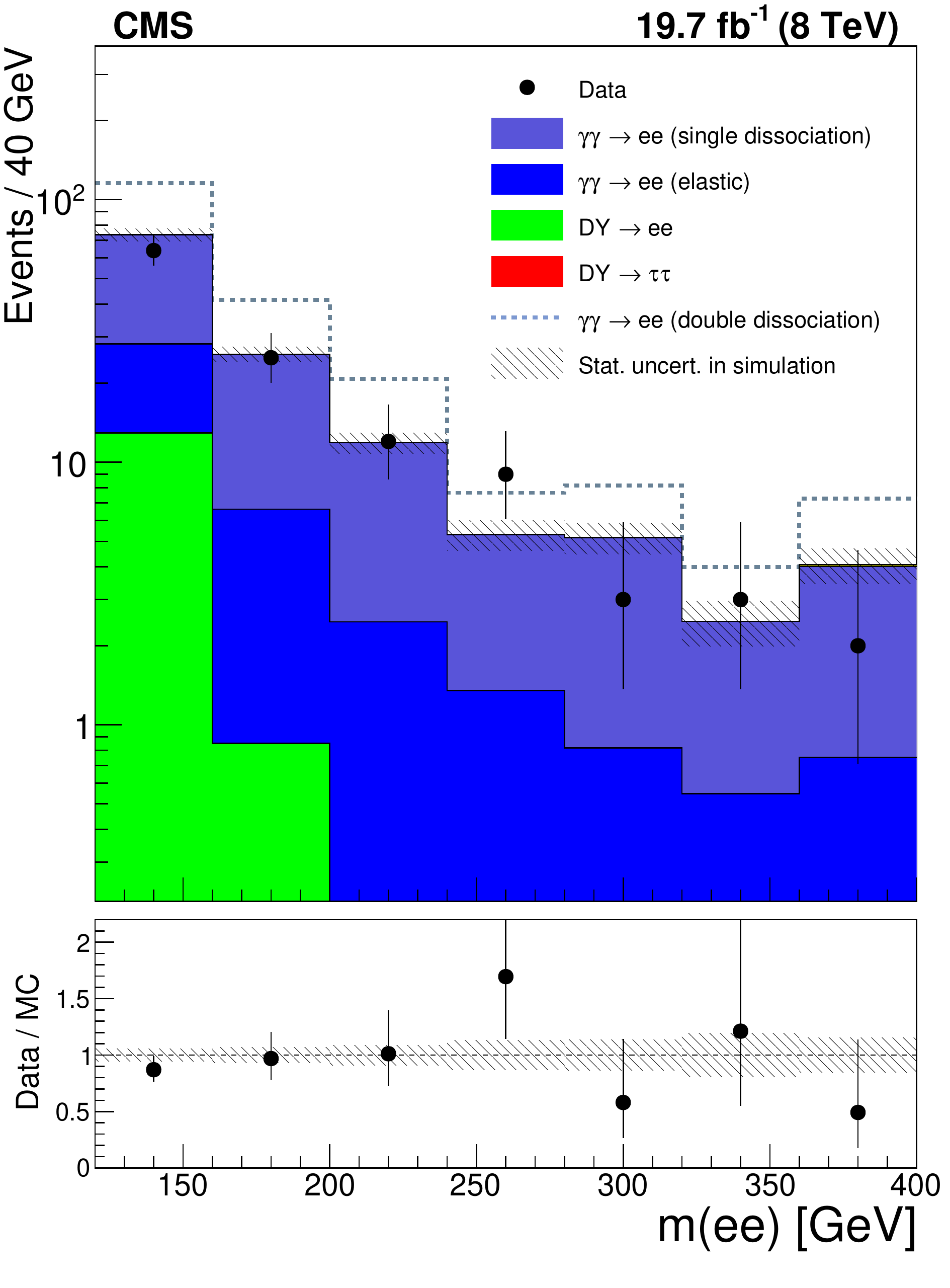}
\caption{Dilepton invariant mass for the $\PGmp \PGmm$ (left) and $\Pep \Pem$ (right) final states in the $\PGg\PGg \to \Plp\Plm$ proton dissociation control region with no additional tracks associated to the dilepton vertex, for data (points with error bars) and simulated samples (histograms, with the  efficiency correction applied to the exclusive samples). The last bin in both plots is an overflow bin and includes all events above the maximum value in the plot. The bottom panels show the data/MC ratio where the denominator includes the sum of all simulated samples except the double-dissociation contribution (shown as the blue dotted line in the top plots).
\label{fig:pdissfactorggllplots}}
\end{figure}

Figure~\ref{fig:pdissfactorggllplots} shows the dilepton invariant mass distribution for events with no additional tracks at the dilepton vertex.
The theoretical double-dissociation contribution (blue dotted line on top of the sum of all other simulated data samples in Fig.~\ref{fig:pdissfactorggllplots}) is much larger than the data,
because the value of the gap survival probability factor is too high in the calculations,
whereas at high dilepton mass the data are consistent with a very low survival probability for this contribution.
For a 100\% gap survival probability in both single- and double-dissociation processes, the scale factor to correct the elastic prediction would be $\mathrm{F} = 7.71 \pm 0.57$, by applying the same procedure described above but using the single- and double-dissociation simulated samples.
However, if the double-dissociation contribution is assumed to have a gap survival probability of 0\% (maximum predicted rescattering), whereas the single-dissociation contribution is assumed to have a gap survival probability of 100\% (no rescattering), the proton dissociation factor estimated from the simulation would be $\mathrm{F} = 4.39 \pm 0.48$. This factor is compatible with that extracted from the data-driven method
described above and is also consistent with the expectation from theory~\cite{HarlandLang:2012qz,Harland-Lang:2016apc} that the single-dissociation contribution has a large gap survival probability while the double-dissociation contribution has a small gap survival probability.

\section{Backgrounds}

\subsection{Inclusive diboson backgrounds}

\begin{figure}[htb]
\centering
\includegraphics[width=.45\textwidth]{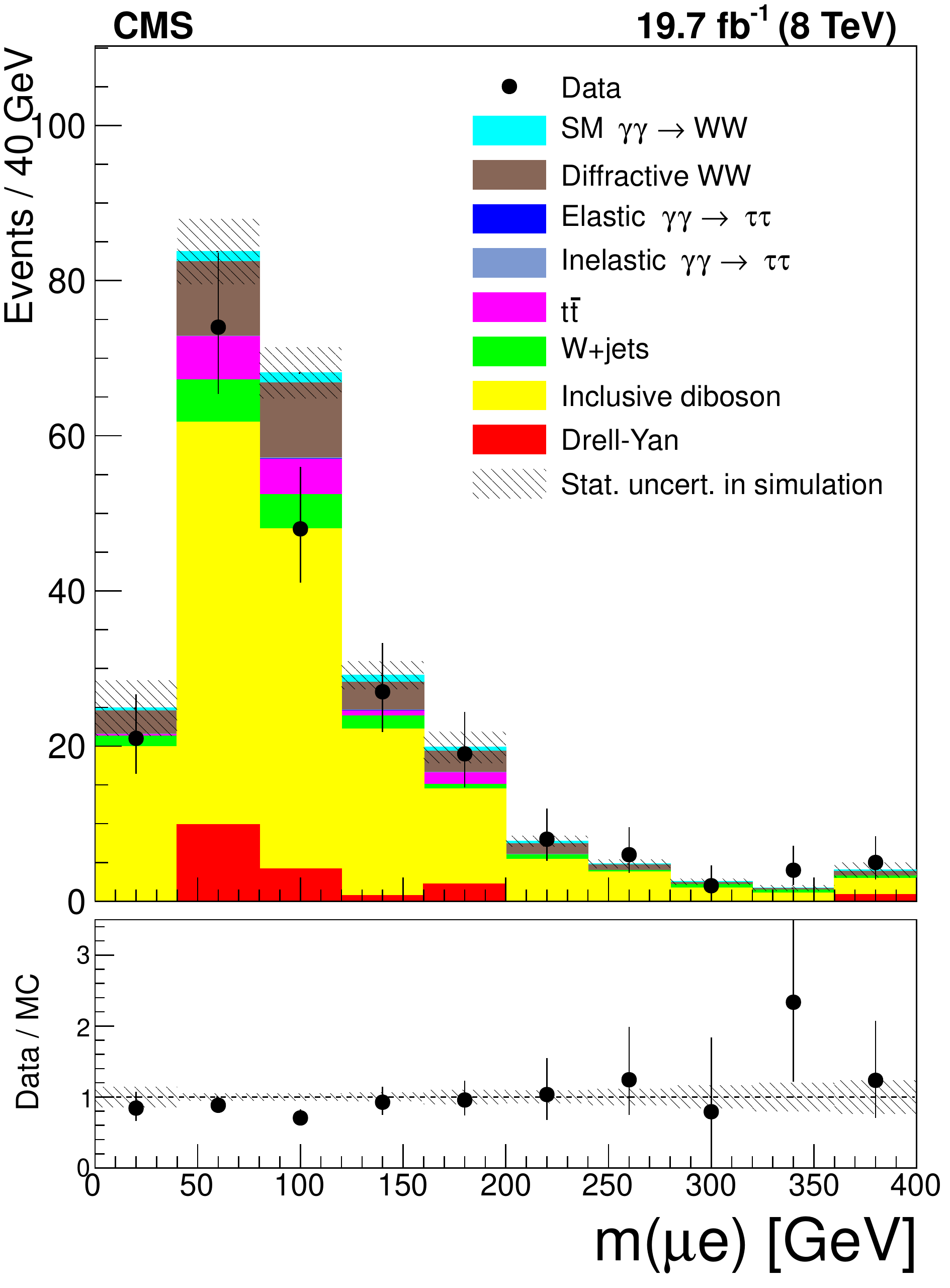}
\includegraphics[width=.45\textwidth]{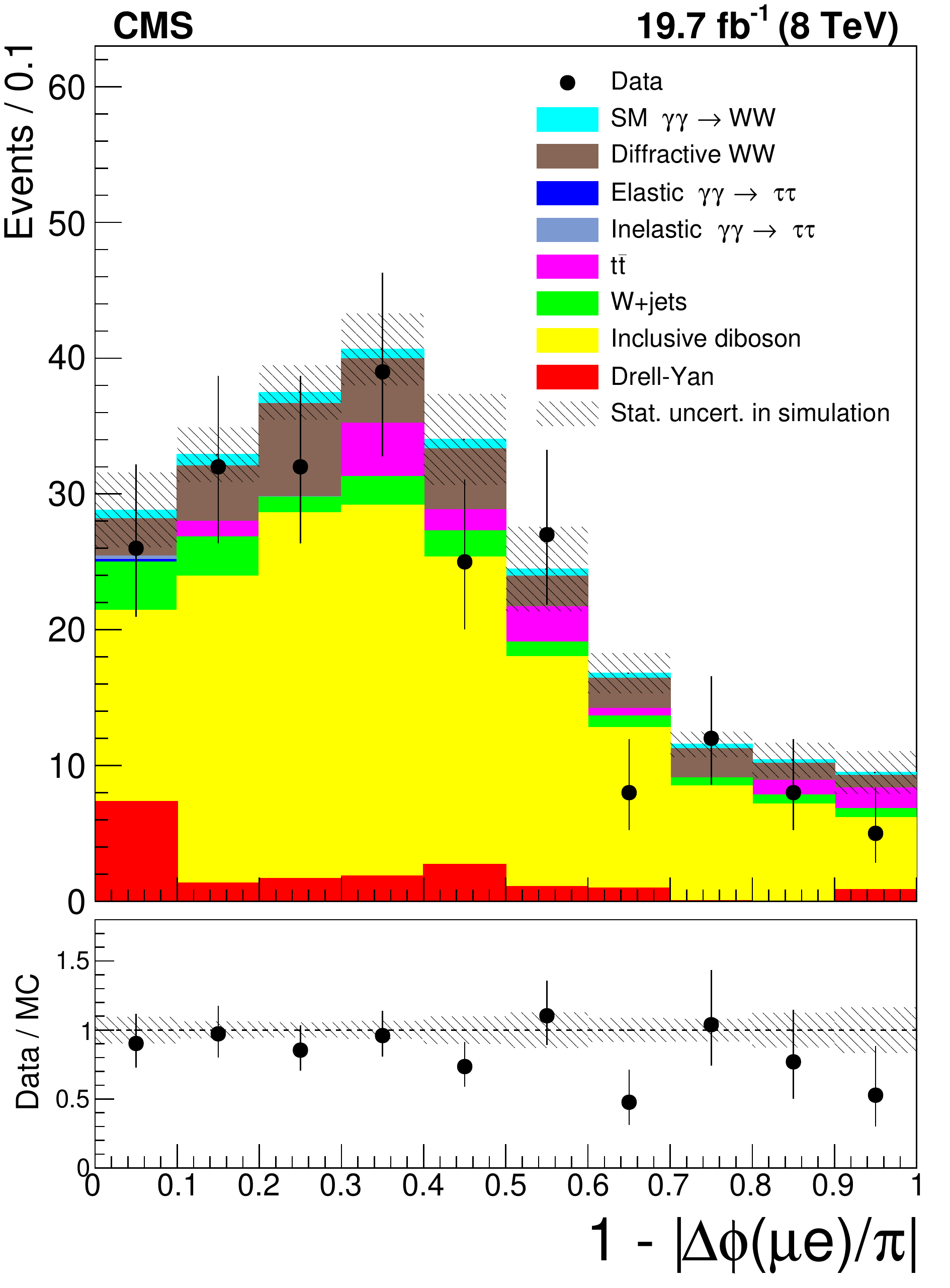}
\caption{Distributions of $\PGm^{\pm}\Pe^{\mp}$ invariant mass (left) and acoplanarity (right) for data (points with error bars) and expected backgrounds (histograms) for $\pt(\PGm^{\pm}\Pe^{\mp}) > 30$\GeV and 1--6 extra tracks (inclusive $\PWp\PWm$ control region). The last bin in the invariant mass plot is an overflow bin and includes all events with $m(\PGm\Pe)> 360$\GeV. The bottom panels show the data/MC
ratio.
\label{fig:inclwwplots}}
\end{figure}

The dominant inclusive diboson backgrounds consist mainly of $\PWp\PWm$ events, with a small contribution from $\PW\Z$ and $\Z\Z$ events. As indicated in
Table~\ref{tab:EMucutflowtable}, the inclusive diboson background is reduced by a factor of more than 300 by vetoing on additional tracks
at the
$\PGm^{\pm}\Pe^{\mp}$ vertex. The remaining backgrounds are studied by selecting electron-muon vertices with $\pt(\PGm^{\pm}\Pe^{\mp}) >30$\GeV,
and 1--6 additional tracks. The event yields and kinematic distributions are compatible with the expectations from simulation, with
$247.0 \pm 8.0\stat$ events expected and 214 events observed in data (Fig.~\ref{fig:inclwwplots}).

The inclusive $\PWp\PWm$ background estimate obtained using \MADGRAPH  for the signal region (no additional tracks
and $\pt(\PGm^{\pm}\Pe^{\mp}) >  30$\GeV) is $2.2 \pm 0.4\stat$  events. The prediction is additionally cross-checked with an estimate based on \PYTHIA, which also describes well the control region with 1--6 extra tracks. This results in an inclusive $\PWp\PWm$ background prediction of $2.5 \pm 0.9\stat$ events, consistent with the default prediction using \MADGRAPH.
The $\PW\Z$ and $\Z\Z$ background estimates are obtained from \MADGRAPH as well, and only contribute $0.1 \pm 0.1\stat$ events in the signal region.

\subsection{W+jets background}

The background due to W+jets production, with one genuine lepton and one misidentified or nonprompt lepton in a jet, is expected to be small. To study this background in data, a control
sample, expected to be dominated by W+jets, is selected with the $\pt(\PGm^{\pm}\Pe^{\mp})>$30\GeV requirement, where at least one of the two leptons has failed the nominal offline identification described in Section~\ref{sec:EventSelection}. The control
sample is expected to contain 78\% W+jets events.
To extract a prediction for the W+jets background contribution in which both leptons pass the nominal lepton identification requirements, we use
the ratio of the number of events in the signal and control regions calculated from simulation and multiply this ratio by the
number of data events in the control region.
The resulting prediction in the signal region is $0.2\pm0.1\stat$ events, approximately 5\% of the total background.

\subsection{Drell--Yan background}

\begin{figure}[htb]
\centering
\includegraphics[width=.45\textwidth]{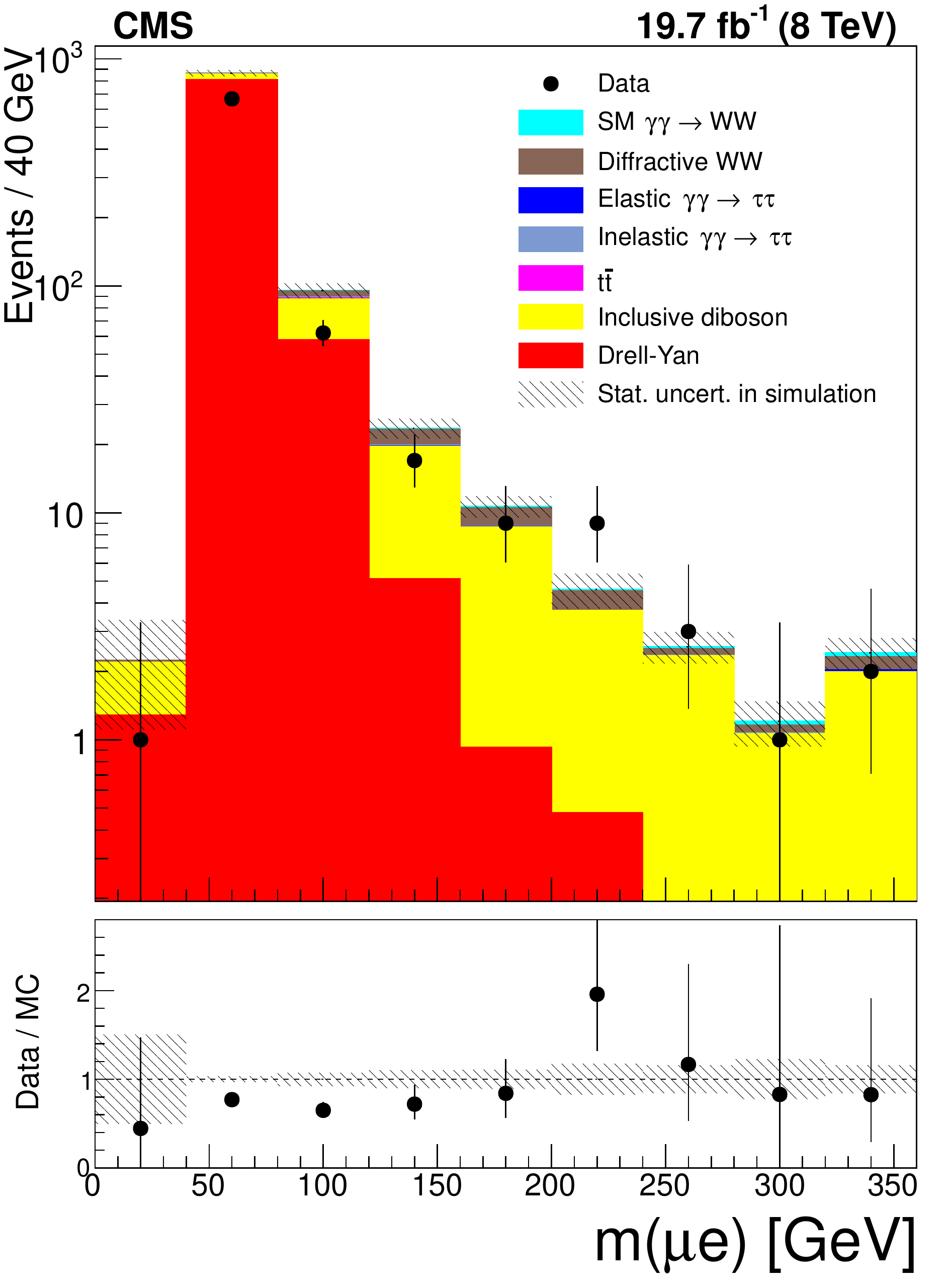}
\includegraphics[width=.45\textwidth]{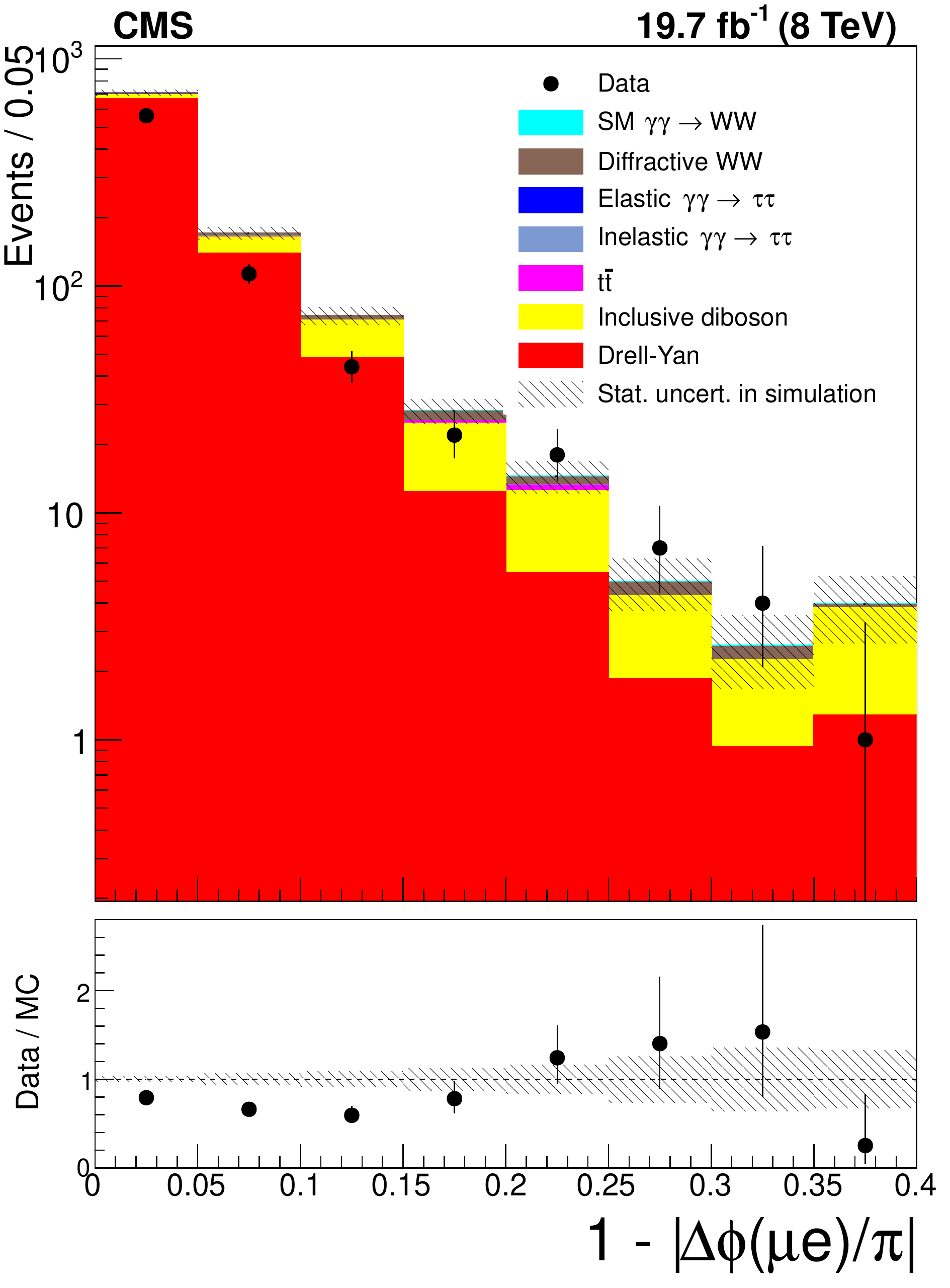}
\caption{
Distributions of $\PGm^{\pm}\Pe^{\mp}$ invariant mass (left) and acoplanarity (right) for data (points with error bars) and expected backgrounds (histograms) for $\pt(\PGm^{\pm}\Pe^{\mp}) < 30$\GeV and 1--6 extra tracks (Drell--Yan $\PGt^{+}\PGt^{-}$ control region). The last bin in both plots is an overflow bin and includes all events above the maximum value in the plot. The bottom panels show the data/MC ratio.
\label{fig:dytautauplots}}
\end{figure}

The background due to DY $\PGt^{+}\PGt^{-}$ production is suppressed by a factor of more than 700 by the requirement of no additional tracks associated with the $\PGm^{\pm}\Pe^{\mp}$ vertex (Table~\ref{tab:EMucutflowtable}). To check the modeling of the DY background contribution, a control region with
$\pt(\PGm^{\pm}\Pe^{\mp})<30$\GeV and 1--6 additional tracks is selected, resulting in a
sample that is expected to contain 87\% DY $\PGt^{+}\PGt^{-}$ events.
We
find an overall deficit in the data with respect to the prediction from simulation,
with 771 events observed and $1008 \pm 27\stat$ events expected.
Figure~\ref{fig:dytautauplots} shows that this deficit appears at low mass and low-acoplanarity where the DY background is expected.
At higher values of the mass and acoplanarity where the inclusive $\PWp\PWm$ contribution is significant, the data agree well with the simulation,
consistent with the behavior observed in the $\PWp\PWm$ control region. The number
of simulated DY events surviving in the signal region after all selections is zero, therefore no rescaling of the DY background is performed based
on the control region yields.

\subsection{The \texorpdfstring{$\PGg\PGg\to\PGt^{+}\PGt^{-}$}{gamma-gamma to tau+ tau-} background}

As $\PGg\PGg\to\PGt^{+}\PGt^{-}$ is produced in both exclusive and quasi-exclusive topologies, it cannot be completely eliminated by requiring
no additional tracks at the $\PGm^{\pm}\Pe^{\mp}$ vertex. The requirement that $\pt(\PGm^{\pm}\Pe^{\mp}) > 30$\GeV, however, combined with the 20\GeV single-lepton
thresholds, reduces this background to
approximately one event in the signal region (Table~\ref{tab:EMucutflowtable}).

\begin{figure}[htb]
\centering
\includegraphics[width=.45\textwidth]{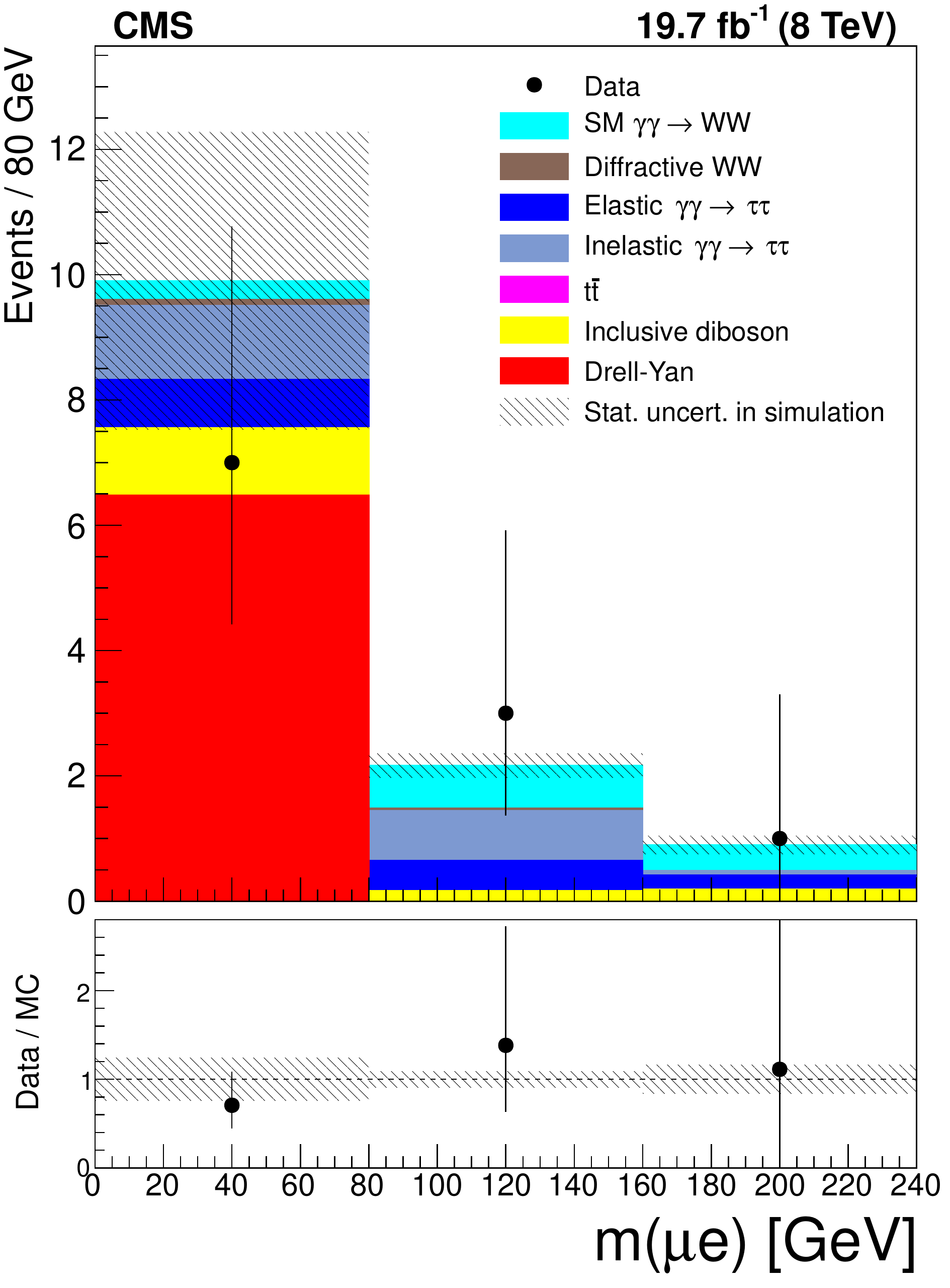}
\includegraphics[width=.45\textwidth]{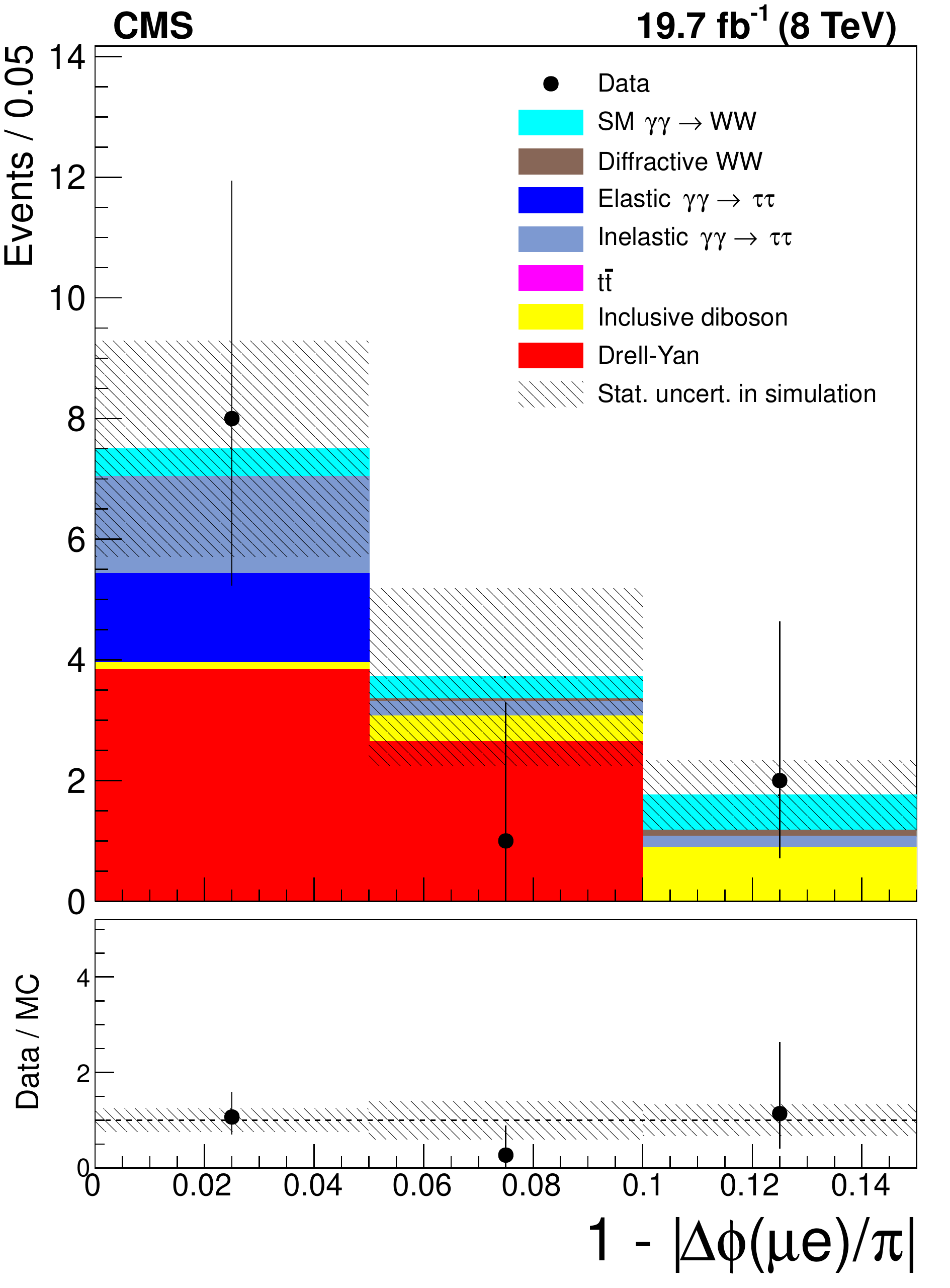}
\caption{
Distributions of $\PGm^{\pm}\Pe^{\mp}$ invariant mass (left) and acoplanarity (right) for data (points with error bars) and expected backgrounds (histograms) for $\pt(\PGm^{\pm}\Pe^{\mp}) < 30$\GeV and no additional tracks ($\PGg\PGg~\to~\PGt^{+}\PGt^{-}$ control region). The last bin in both plots is an overflow bin and includes all events above the maximum value in the plot. The bottom panels show the data/MC ratio.
\label{fig:gamgamtautauplots}}
\end{figure}

A control sample enriched in $\PGg\PGg\to\PGt^{+}\PGt^{-}$ events is
selected by requiring an electron-muon vertex with no additional associated tracks,
and $\pt(\PGm^{\pm}\Pe^{\mp}) < 30$\GeV. In data, 11 events are observed, compared to a prediction of $12.9 \pm 2.5$,
including $3.4 \pm 0.5$ events expected from $\PGg\PGg\to\PGt^{+}\PGt^{-}$ production. The kinematic distributions
are in good agreement with the predicted sum of $\PGg\PGg\to\PGt^{+}\PGt^{-}$ and other
backgrounds (Fig.~\ref{fig:gamgamtautauplots}).

\subsection{Summary of backgrounds}

The number of expected signal and background events at each stage of the selection is shown in Table~\ref{tab:EMucutflowtable}. As described in Section~\ref{sec:DatasetsMC},
the diffractive $\PWp\PWm$ background is estimated from simulation, assuming the maximal gap survival probability of 100\%. By assuming a smaller survival probability
the total background prediction would decrease by at most 0.1 events, which is less than 3\% of the total background and less than 2\% of the expected SM signal.
The ``Other backgrounds'' category includes the contributions of t$\bar{\mathrm{t}}$, W+jets, electroweak $\PWp\PWm\PQq\PQq$, diffractive $\PWp\PWm$, and jets. The total expected background is $3.9 \pm 0.6$ events, with the largest contribution coming from inclusive $\PWp\PWm$ production. The
expected SM signal is $5.3 \pm 0.7$ events.

As a final check for potential mismodeled backgrounds, we examine same-charge $\PGm^{\pm}\Pe^{\pm}$ events. In the control region with 1--6 extra tracks and
$\pt(\PGm^{\pm}\Pe^{\pm}) > 30$\GeV, 28 such events are observed, with track multiplicity and invariant mass distributions consistent with the simulation, which
predicts $20.6\pm2.1$ events. In the signal-like region with no additional tracks and $\pt(\PGm^{\pm}\Pe^{\pm}) > 30$\GeV, no same-charge events are observed in the data,
consistent with the prediction of 0.12 background events from simulation.

\begin{table}[htb]
\centering
\topcaption{Number of expected signal and background events in simulation passing each selection step, normalized to an integrated luminosity of 19.7\fbinv. The preselection includes events with an opposite-charge muon and electron associated with the same vertex, each with $\pt>20$\GeV and $\abs{\eta}<2.4$, and $<$16 additional tracks at the vertex. Uncertainties are statistical only.}
\resizebox{\textwidth}{!} {
\begin{tabular}{l|c|c|c|cccccc}
\hline
Selection step & Data & Exclusive & Total & Inclusive &  Drell-Yan & $\PGg\PGg \to \PGt\PGt$  & Other\\
               & & $\PGg\PGg \to  \PW\PW$  & background & diboson &  &  & backgrounds\\
\hline
Trigger and Preselection & 19406 & 26.9$\pm$0.2 & 22180$\pm$1890 & 1546$\pm$15 &  7093$\pm$75 &18.1$\pm$0.8 & 13520$\pm$1890\\
$ m(\PGm^{\pm}\Pe^{\mp})>  20$\GeV & 18466 & 26.6$\pm$0.2 & 21590$\pm$1850 & 1507$\pm$15 & 7065$\pm$75 & 18.1$\pm$0.8 & 13000$\pm$1850\\
Muon and electron identification& 6541 & 22.5$\pm$0.2 & 6640$\pm$93 & 1306$\pm$11 &  4219$\pm$58 & 12.6$\pm$0.7 & 1102$\pm$72\\
$ \PGm^{\pm}\Pe^{\mp} $ vertex with no add. tracks & 24 & 6.7$\pm$0.2 & 15.2$\pm$2.5 & 3.7$\pm$0.7 & 6.5$\pm$2.3 & 4.3$\pm$0.5 &  0.7$\pm$0.1\\
$ \pt(\PGm^{\pm}\Pe^{\mp})> 30$\GeV & 13 & 5.3$\pm$0.1 & 3.9$\pm$0.5 & 2.3$\pm$0.4 &  0.1$\pm$0.1  & 0.9$\pm$0.2 & 0.6$\pm$0.1\\
\hline
\end{tabular}
}

\label{tab:EMucutflowtable}
\end{table}

\section{Systematic uncertainties}

We consider systematic uncertainties related to the integrated luminosity, the lepton trigger and selection efficiency,
the efficiency of the additional track veto, and the uncertainty in the proton dissociation contribution.

The integrated luminosity uncertainty for the 8\TeV data set used in this measurement is estimated to be 2.6\%~\cite{CMS-PAS-LUM-13-001}. The trigger and
lepton identification efficiencies are corrected for differences between data and simulation using control samples of
$\Z\to\Plp\Plm$ events.
The systematic uncertainty is estimated from the statistical uncertainty associated with the correction applied,
resulting in an uncertainty of 2.4\% in the signal efficiency.

The correction for the efficiency of the additional track veto is obtained from the control samples of elastic-enriched
$\PGg\PGg\to\Plp\Plm$ events, as described in Section~\ref{sec:ControlSamplesCorrections}.
Since the correction factors obtained in the
$\PGmp\PGmm$ and $\Pep\Pem$ channels are consistent, they are combined to obtain the final correction factor.
The systematic uncertainty is estimated from the statistical uncertainty associated with the correction applied,
resulting in an overall uncertainty of 5\% in the signal efficiency.

The normalization factor for the proton dissociation contribution to the signal is obtained from high-mass $\PGg\PGg\to\Plp\Plm$ events in data as explained in Section 6. The statistical uncertainty in this factor is 9.2\%, based on the
combination of the $\PGmp\PGmm$ and $\Pep\Pem$ channels. An additional effect of 5.0\%
must be included to describe the difference between
the matrix element prediction of \textsc{lpair} used in the method described in Section~\ref{sec:ControlSamplesCorrections}, and the equivalent photon approximation used to generate signal events.
Adding in quadrature these contributions results in an overall systematic uncertainty of 10.5\% related to the proton dissociation contribution.
It is also checked that the proton dissociation factor does not vary as a function of the dilepton invariant mass threshold, between 100--400\GeV.

The full list of systematic uncertainties for the signal efficiency is shown in Table~\ref{tab:signalsyst}. The overall systematic uncertainty assigned
to the signal is 12.2\%. The systematic uncertainties considered for the background prediction include the limited statistics of the relevant simulation or
data control samples, integrated luminosity, trigger efficiency, and lepton identification efficiency. In addition, an uncertainty
of $\pm$0.24 events in the electroweak $\PWp\PWm$ background contribution is included, corresponding to the diffe\-rence between the background predictions of
the \MADGRAPH and \PHANTOM generators.

\begin{table}[htb]
\begin{center}
\topcaption{Summary of systematic uncertainties affecting the signal.}
\begin{tabular}{lr}
\hline
                                              & \multicolumn{1}{c}{Uncertainty} \\
\hline
Proton dissociation factor                     & 10.5\% \\
Efficiency correction for no add. tracks               & 5.0\%   \\
Trigger and lepton identification                     & 2.4\%  \\
Integrated luminosity                                    & 2.6\% \\ \hline
Total                                         & 12.2\% \\
\hline
\end{tabular}
\label{tab:signalsyst}
\end{center}
\end{table}

\section{Results}

\begin{figure}[htb]
\centering
\includegraphics[width=.48\textwidth]{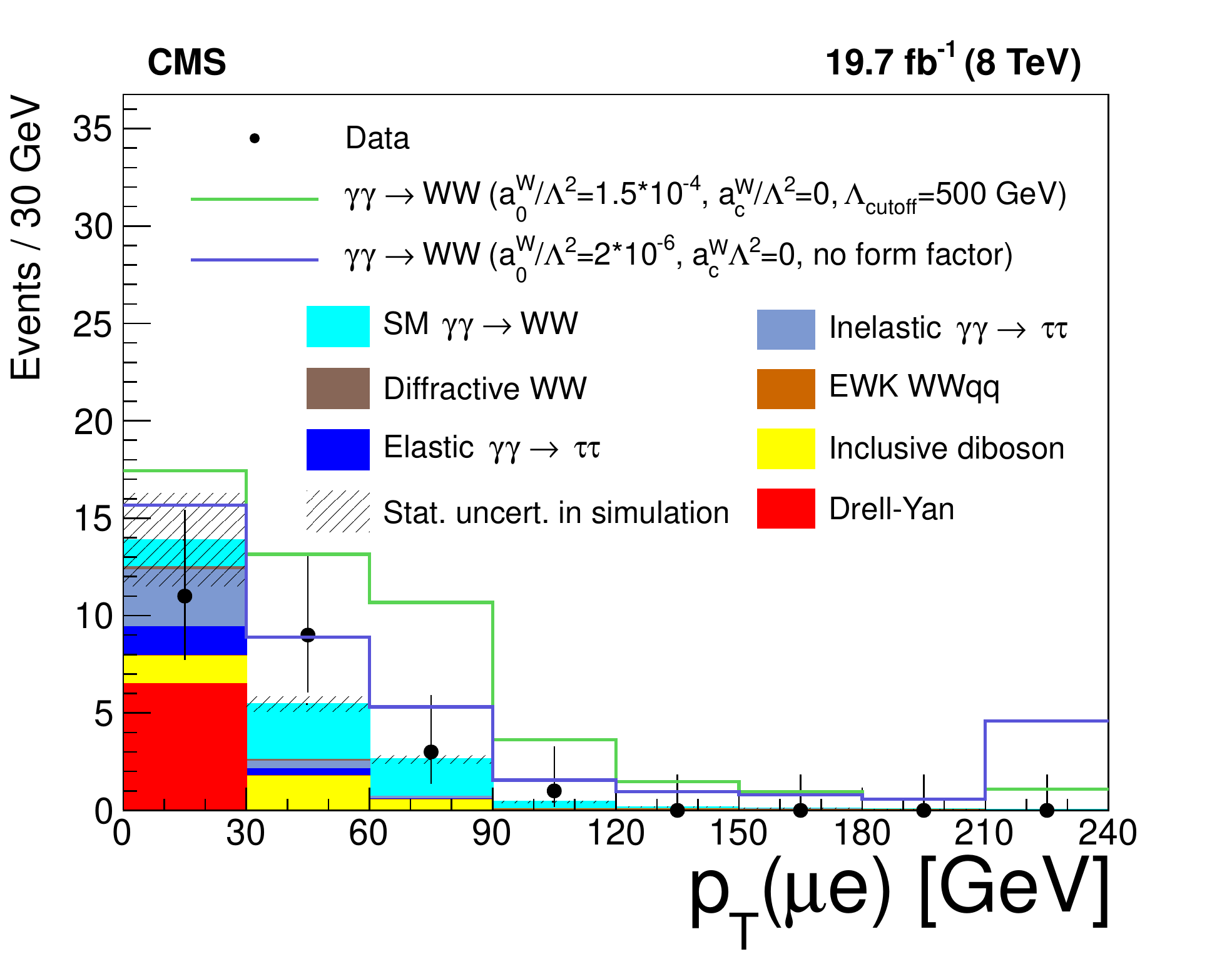}
\includegraphics[width=.48\textwidth]{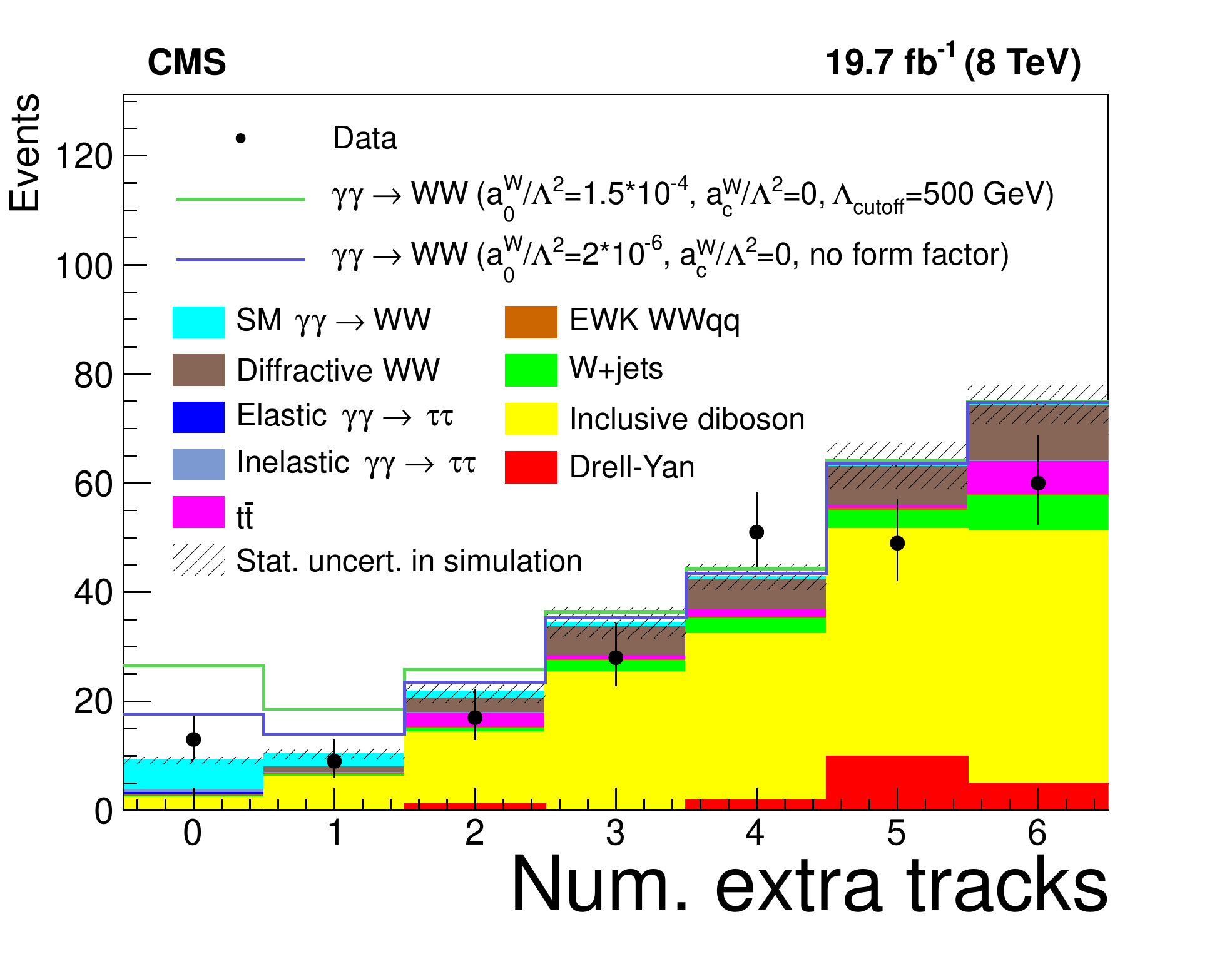}
\caption{Distributions of muon-electron transverse momentum for events with zero associated tracks (left), and extra-tracks multiplicity for events with $\pt(\PGm^{\pm}\Pe^{\mp}) > 30$\GeV (right). The data are shown by the points with error bars; the histograms indicate the expected SM signal and backgrounds.  Two representative values for anomalous couplings are shown stacked on top of the backgrounds.
The last bin in the $\pt(\PGm^{\pm}\Pe^{\mp})$ distribution
is an overflow bin and includes all events with $\pt(\PGm^{\pm}\Pe^{\mp}) > 210$\GeV.
\label{fig:gamgamwwnminusoneplots}}
\end{figure}

\begin{figure}[htb]
\centering
\includegraphics[width=.45\textwidth]{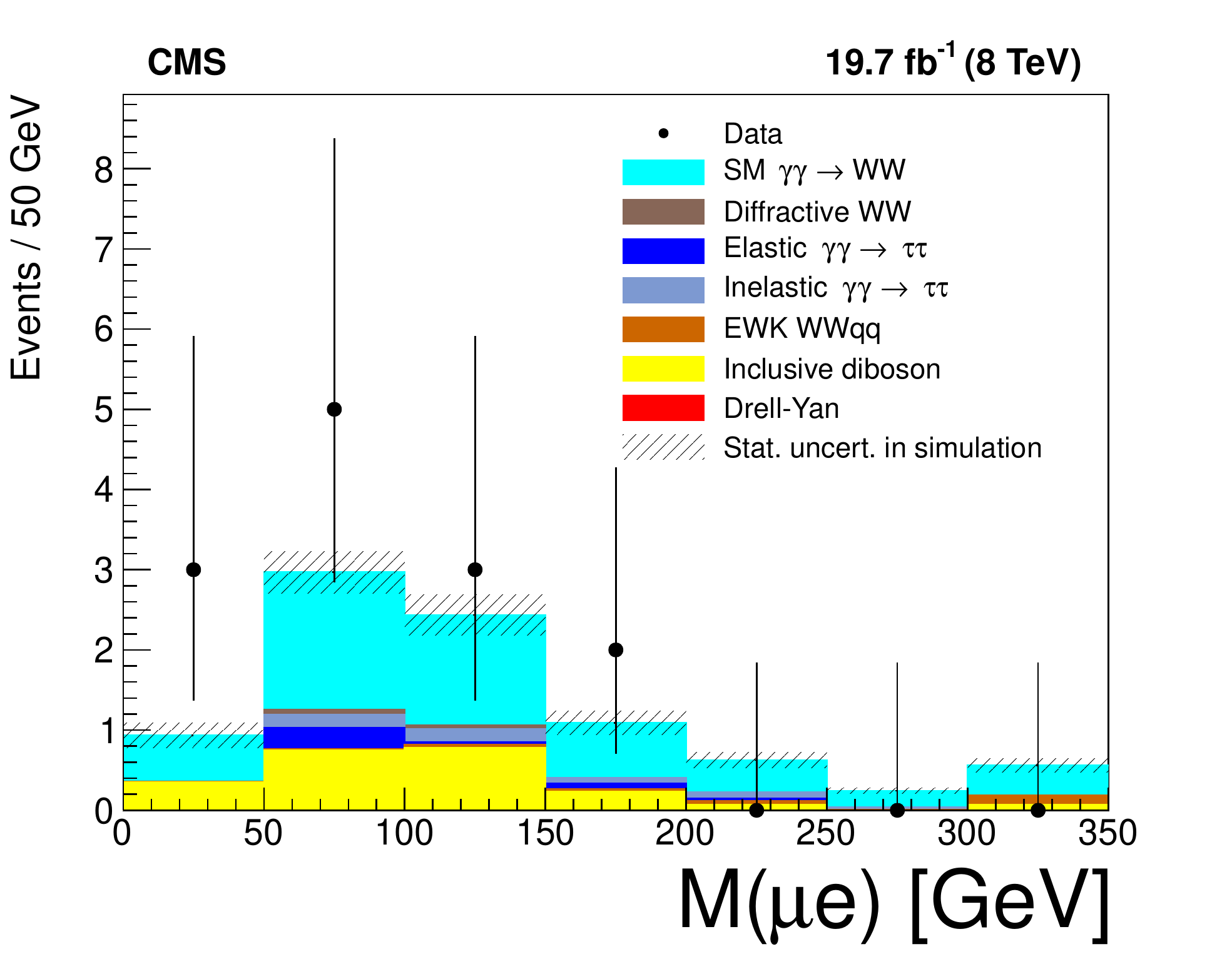}
\includegraphics[width=.45\textwidth]{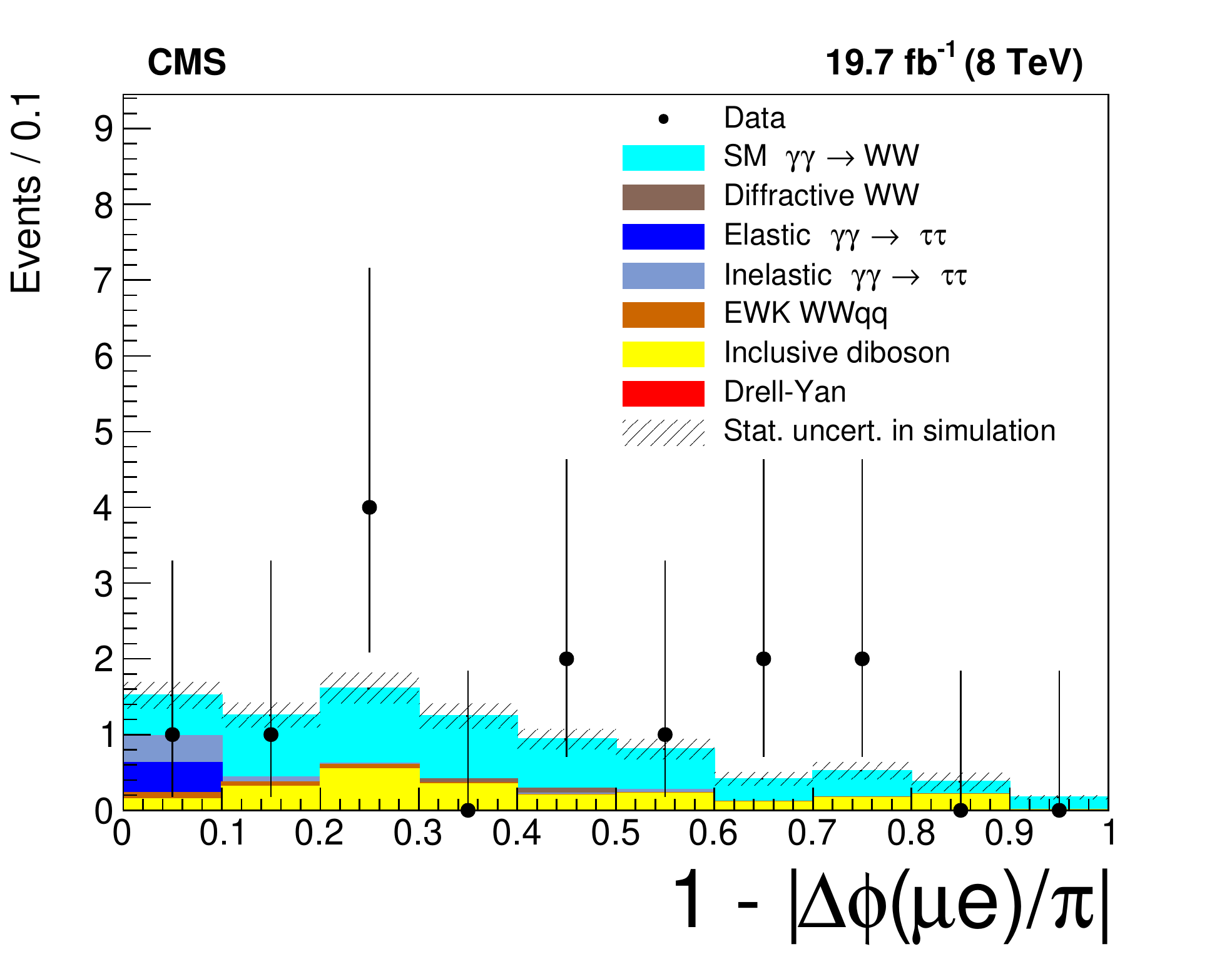}
\includegraphics[width=.45\textwidth]{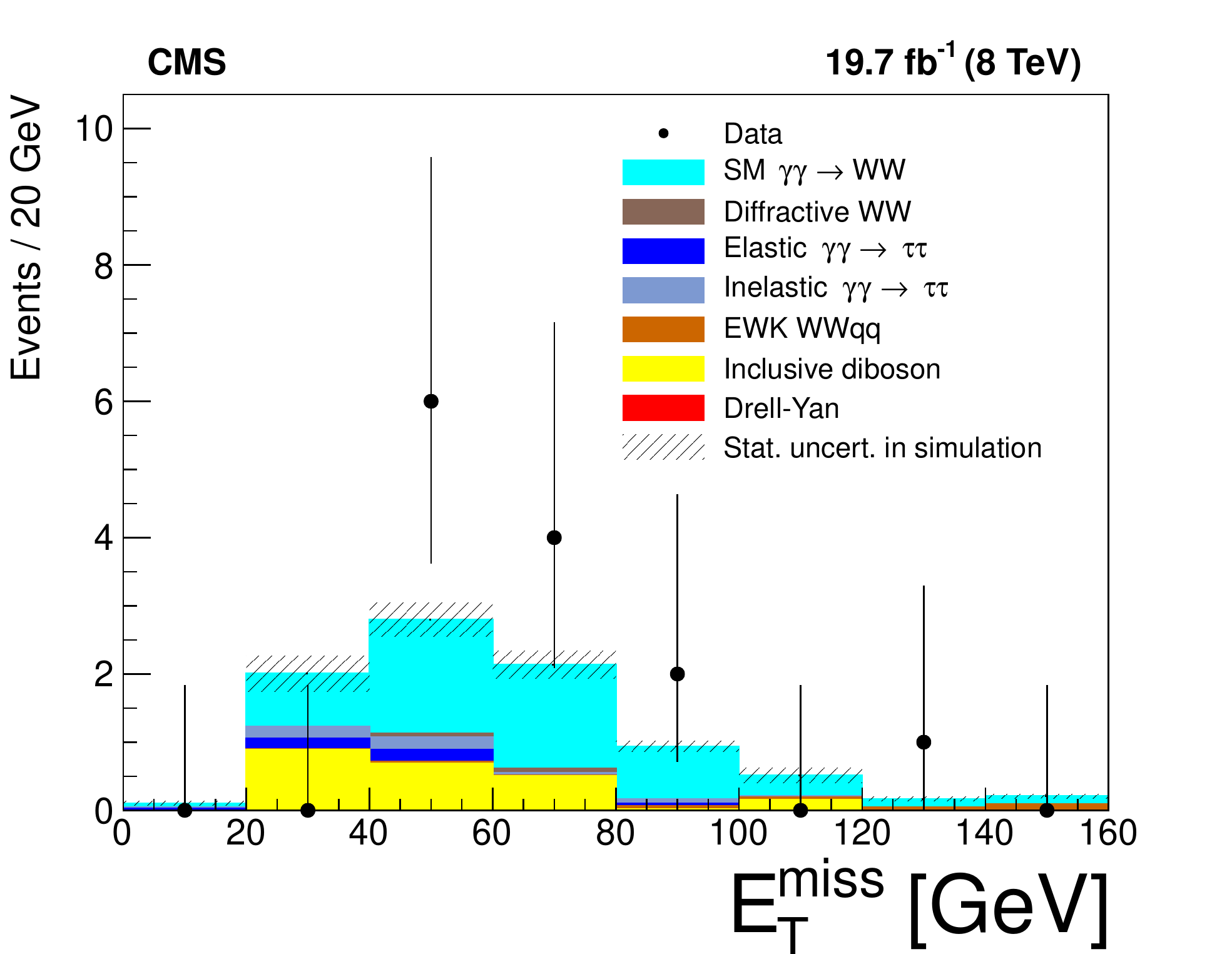}
\caption{Muon-electron invariant mass (top left), acoplanarity (top right), and missing transverse energy (bottom) in
the $\PGg\PGg~\to~\PWp\PWm$ signal region. The data are shown by the points with error bars; the histograms indicate the expected SM
signal and backgrounds.
The last bin in the invariant mass and missing transverse energy plots is an overflow bin and
includes also all events above the maximum value in the plot.
\label{fig:gamgamwwsignalregionplots}}
\end{figure}

The total expected signal from standard model exclusive or quasi-exclusive $\PGg\PGg\to\PWp\PWm$ production in the 8\TeV data set is $5.3 \pm 0.7$ events, with an expected background of $3.9 \pm 0.6$ events. This corresponds to a mean
expected signal significance of $2.1 \sigma$. Figure~\ref{fig:gamgamwwnminusoneplots} shows the $\pt(\PGm^{\pm}\Pe^{\mp})$
and extra-tracks multiplicity distributions for events passing all other selection requirements. In the signal region with no additional tracks and $\pt(\PGm^{\pm}\Pe^{\mp}) > 30$\GeV, 13 events are observed in the data that
pass all the selection criteria. The properties of the selected events, including the $\PGm^{\pm}\Pe^{\mp}$ invariant mass, acoplanarity, and
missing transverse energy ($\MET$), are consistent with the SM signal plus background prediction (Fig.~\ref{fig:gamgamwwsignalregionplots}).

The observed significance above the background-only hypothesis in the 8\TeV data, including systematic uncertainties, is $3.2 \sigma$. In the 7\TeV data,
two events were observed in the signal region, with an expected background of $0.84 \pm 0.15$ events, corresponding to an observed (expected) significance
of 0.8$\sigma$ (1.8$\sigma$). We combine the 7 and 8\TeV results, treating all systematic uncertainties as fully uncorrelated between the two measurements, with
the exception of the 5\% uncertainty from the use of the equivalent photon approximation in the generation of signal samples, which is treated as fully correlated between the two analyses.
The resulting observed (expected) significance for the 7 and 8\TeV combination is 3.4$\sigma$ (2.8$\sigma$), constituting evidence for $\PGg\PGg \to \PWp\PWm$ production
in proton-proton collisions at the LHC.

\subsection{Cross section measurement}

Interpreting the 8\TeV results as a cross section multiplied by the branching fraction to $\PGm^{\pm}\Pe^{\mp}$ final states, corrected
for all experimental efficiencies and extrapolated to the full phase space, yields:
\begin{equation*}
\sigma(\Pp\Pp \to \Pp^{(*)}\PWp\PWm\Pp^{(*)} \to \Pp^{(*)}\PGm^{\pm}\Pe^{\mp}\Pp^{(*)}) = 10.8^{+5.1}_{-4.1}\unit{fb}.
\end{equation*}

The SM prediction is $6.2 \pm 0.5\unit{fb}$, with the elastic component calculated with \MADGRAPH
and then rescaled by the proton dissociation factor.
The uncertainty on the SM prediction reflects the uncertainty in the proton dissociation contribution
to the signal. The acceptance for the SM signal calculated from the simulation is 57.8 $\pm$ 0.9\%.

The corresponding 95\% confidence level (CL) upper limit obtained from the 7\TeV data was~\cite{Chatrchyan:2013akv}:
\begin{equation*}
\sigma(\Pp\Pp \to \Pp^{(*)}\PWp\PWm\Pp^{(*)} \to \Pp^{(*)}\PGm^{\pm}\Pe^{\mp}\Pp^{(*)}) < 10.6\unit{fb},
\end{equation*}
with a central value of $2.2^{+3.3}_{-2.0}\unit{fb}$. The corresponding SM prediction at 7\TeV is $4.0 \pm 0.7\unit{fb}$, with the uncertainty reflecting that of the proton dissociation contribution to the signal.

\subsection{Anomalous couplings}

We use the dilepton transverse momentum $\pt(\PGm^{\pm}\Pe^{\mp})$ (Fig.~\ref{fig:gamgamwwnminusoneplots}, left) as a discriminating variable to extract limits on AQGCs.
Two bins, with boundaries $\pt(\PGm^{\pm}\Pe^{\mp}) =30$--130\GeV and $\pt(\PGm^{\pm}\Pe^{\mp}) > 130$\GeV, are used in the limit setting procedure for the 8\TeV
analysis. The bin boundaries are chosen such that the a priori expectation for SM $\PGg\PGg \to \PWp\PWm$ in the highest bin is $\sim$0.1 events,
with other backgrounds, predominantly electroweak $\PWp\PWm$ production, contributing an additional $\sim$0.1 events. In the 7\TeV analysis~\cite{Chatrchyan:2013akv} a single bin
with $\pt(\PGm^{\pm}\Pe^{\mp}) > 100$\GeV was used, also chosen such that the a priori expectation for SM $\PGg\PGg \to \PWp\PWm$ is $\sim$0.1 events.

In both the 7 and 8\TeV analyses, and in the combination, the Feldman--Cousins pres\-cription~\cite{Feldman:1997qc} is used to derive limits.
In the 7\TeV analysis, where the number of expected and observed events was near zero, the inclusion of systematic uncertainties
in the background estimate resulted in a shortening of the 95\% confidence interval. Therefore a conservative procedure of integrating the
systematic uncertainties out,
reproducing the method advocated by Cousins and Highland~\cite{Cousins:1991qz}, was used. In the 8\TeV analysis and in the 7+8\TeV combination, no such effect is observed,
therefore the systematic uncertainties are included as log-normal nuisance parameters in the limit calculation. As in the case of the combined significance calculations, the systematic
uncertainties are treated as uncorrelated between the two data sets, except for the EPA uncertainty, which is fully correlated.

Table~\ref{tab:FinalLimitTable} summarizes all of the limits on the dimension-6 and dimension-8 AQGC parameters obtained from the
7 and 8\TeV $\PGg\PGg \to \PWp\PWm$ data separately, and from the combination of the two. The 7\TeV dimension-6 results are taken from Ref.~\cite{Chatrchyan:2013akv},
and translated into the dimension-8 formalism as described in Section~\ref{tab:AQGCTheory}, using Eq.~(1). For these limits all parameters except the one shown are fixed to zero (the value expected in the standard model).
The
8\TeV results are an improvement over previously published values
with $\Lambda_{\text{cutoff}}=500\GeV$~\cite{Chatrchyan:2013akv,Abazov:2013opa,Aad:2015uqa}, of
which  the CMS 7\TeV limits of $1.5 \times 10^{-4}\GeV^{-2}$ and $5 \times 10^{-4}\GeV^{-2}$ on $a^{\PW}_{0}/\Lambda^{2}$ and $a^{\PW}_{C}/\Lambda^{2}$, respectively,
are the most stringent. These limits are also approximately two orders of magnitude more stringent than those obtained at
LEP~\cite{Heister:2004yd,Abbiendi:2004bf,Abbiendi:2003jh,Abbiendi:1999aa,Abdallah:2003xn,Achard:2002iz,Achard:2001eg}, where unitarity was
approximately preserved without form factors, due to the lower $\sqrt{s}$ of e$^{+}$e$^{-}$ collisions. By combining the 7 and 8\TeV data sets, we find upper limits at 95\% CL that are
$\sim$10\% more restrictive than the 8\TeV results alone, for the case of a dipole form factor with $\Lambda_{\text{cutoff}} = 500$\GeV.

With no form factor corrections, there is nothing to prevent the rapidly increasing cross section from violating unitarity at high energies in the theory. We also obtain exclusion results in this scenario, listed
in Table~\ref{tab:FinalLimitTable}, for comparison with other unitarity-violating limits on the same operators~\cite{Chatrchyan:2013akv,Abazov:2013opa,Chatrchyan:2014bza,Khachatryan:2014sta,Aad:2015uqa}.
In this case the high-energy behavior leads to a larger improvement when comparing the 8\TeV to the 7\TeV results in the $\PGg\PGg \to \PWp\PWm$ channel. The dominance of
the 8\TeV results in the unitarity-violating limits also results in only a very small improvement when they are combined with the 7\TeV limits.

\begin{table}[htb]
\topcaption{\label{tab:FinalLimitTable}Summary of all 95\% CL AQGC limits derived from the measured $\pt(\PGm\Pe)$ distributions in the $\PGg\PGg \to \PWp\PWm$ signal region production in CMS at 7 and 8\TeV. The second column lists the 7\TeV limits on dimension-6 operators taken from Ref.~\cite{Chatrchyan:2013akv}, as well as their conversion to dimension-8 operators at 7\TeV. The third column contains the 8\TeV results described in this paper. The final column shows the combined 7 and 8\TeV limits. }
\renewcommand*{\arraystretch}{1.15}
\resizebox{\textwidth}{!} {
\begin{tabular}{l|c|c|c}
\hline
Dimension-6 AQGC parameter & 7\TeV (${\times}10^{-4}\GeV^{-2}$) & 8\TeV (${\times}10^{-4}\GeV^{-2}$) & 7+8\TeV (${\times}10^{-4}\GeV^{-2}$) \\
\hline
$ a^{\PW}_{0}/\Lambda^{2} (\Lambda_{\text{cutoff}}=500\GeV) $ & $-1.5 < a^{\PW}_{0}/\Lambda^{2} < 1.5$ & $-1.1 < a^{\PW}_{0}/\Lambda^{2} < 1.0$ & $-0.9 < a^{\PW}_{0}/\Lambda^{2} < 0.9$ \\
$ a^{\PW}_{C}/\Lambda^{2} (\Lambda_{\text{cutoff}}=500\GeV) $ & $-5 < a^{\PW}_{C}/\Lambda^{2} < 5$ & $-4.2 < a^{\PW}_{C}/\Lambda^{2} < 3.4$ & $-3.6 < a^{\PW}_{C}/\Lambda^{2} < 3.0$ \\
\hline
Dimension-8 AQGC parameter & 7\TeV (${\times}10^{-10}\GeV^{-4}$) & 8\TeV (${\times}10^{-10}\GeV^{-4}$) & 7+8\TeV (${\times}10^{-10}\GeV^{-4}$) \\
\hline
$  f_{M,0}/\Lambda^{4} (\Lambda_{\text{cutoff}}=500\GeV) $ & $ -5.7 < f_{M,0}/\Lambda^{4} < 5.7 $ & $-3.8 < f_{M,0}/\Lambda^{4} < 4.2$ & $-3.4 < f_{M,0}/\Lambda^{4} < 3.4$ \\
$  f_{M,1}/\Lambda^{4} (\Lambda_{\text{cutoff}}=500\GeV) $ & $ -19 < f_{M,1}/\Lambda^{4} < 19 $ & $-16 < f_{M,1}/\Lambda^{4} < 13$ & $-14 < f_{M,1}/\Lambda^{4} < 12$ \\
$  f_{M,2}/\Lambda^{4} (\Lambda_{\text{cutoff}}=500\GeV) $ & $ -2.8 < f_{M,2}/\Lambda^{4} < 2.8 $ & $-1.9 < f_{M,2}/\Lambda^{4} < 2.1$ & $-1.9 < f_{M,2}/\Lambda^{4} < 1.9$ \\
$  f_{M,3}/\Lambda^{4} (\Lambda_{\text{cutoff}}=500\GeV) $ & $ -9.5 < f_{M,3}/\Lambda^{4} < 9.5 $ & $-8.0 < f_{M,3}/\Lambda^{4} < 6.5$ & $-6.8 < f_{M,3}/\Lambda^{4} < 5.7$ \\
\hline
\hline
Dimension-6 AQGC parameter & 7\TeV (${\times}10^{-6}\GeV^{-2}$) & 8\TeV (${\times}10^{-6}\GeV^{-2}$) & 7+8\TeV (${\times}10^{-6}\GeV^{-2}$) \\
\hline
$ a^{\PW}_{0}/\Lambda^{2} (\text{no form factor}) $ & $-4 < a^{\PW}_{0}/\Lambda^{2} < 4$ & $-1.2 < a^{\PW}_{0}/\Lambda^{2} < 1.2$ & $-1.1 < a^{\PW}_{0}/\Lambda^{2} < 1.1$ \\
$ a^{\PW}_{C}/\Lambda^{2} (\text{no form factor}) $ & $-15 < a^{\PW}_{C}/\Lambda^{2} < 15$ & $-4.4 < a^{\PW}_{C}/\Lambda^{2} < 4.4$ & $-4.1 < a^{\PW}_{C}/\Lambda^{2} < 4.1$ \\
\hline
Dimension-8 AQGC parameter & 7\TeV (${\times}10^{-12}\GeV^{-4}$) & 8\TeV (${\times}10^{-12}\GeV^{-4}$) & 7+8\TeV (${\times}10^{-12}\GeV^{-4}$) \\
\hline
$  f_{M,0}/\Lambda^{4} (\text{no form factor}) $ & $-15 < f_{M,0}/\Lambda^{4} < 15$ & $-4.6 < f_{M,0}/\Lambda^{4} < 4.6$ & $-4.2 < f_{M,0}/\Lambda^{4} < 4.2$ \\
$  f_{M,1}/\Lambda^{4} (\text{no form factor}) $ & $-57 < f_{M,1}/\Lambda^{4} < 57$ & $-17 < f_{M,1}/\Lambda^{4} < 17$   & $-16 < f_{M,1}/\Lambda^{4} < 16$\\
$  f_{M,2}/\Lambda^{4} (\text{no form factor}) $ & $-7.6 < f_{M,2}/\Lambda^{4} < 7.6$ & $-2.3 < f_{M,2}/\Lambda^{4} < 2.3$ & $-2.1 < f_{M,2}/\Lambda^{4} < 2.1$ \\
$  f_{M,3}/\Lambda^{4} (\text{no form factor}) $ & $-28 < f_{M,3}/\Lambda^{4} < 28$ & $-8.4 < f_{M,3}/\Lambda^{4} < 8.4$ & $-7.8 < f_{M,3}/\Lambda^{4} < 7.8$ \\
\hline
\end{tabular}}
\end{table}

\begin{figure}
\centering
\includegraphics[width=.70\textwidth]{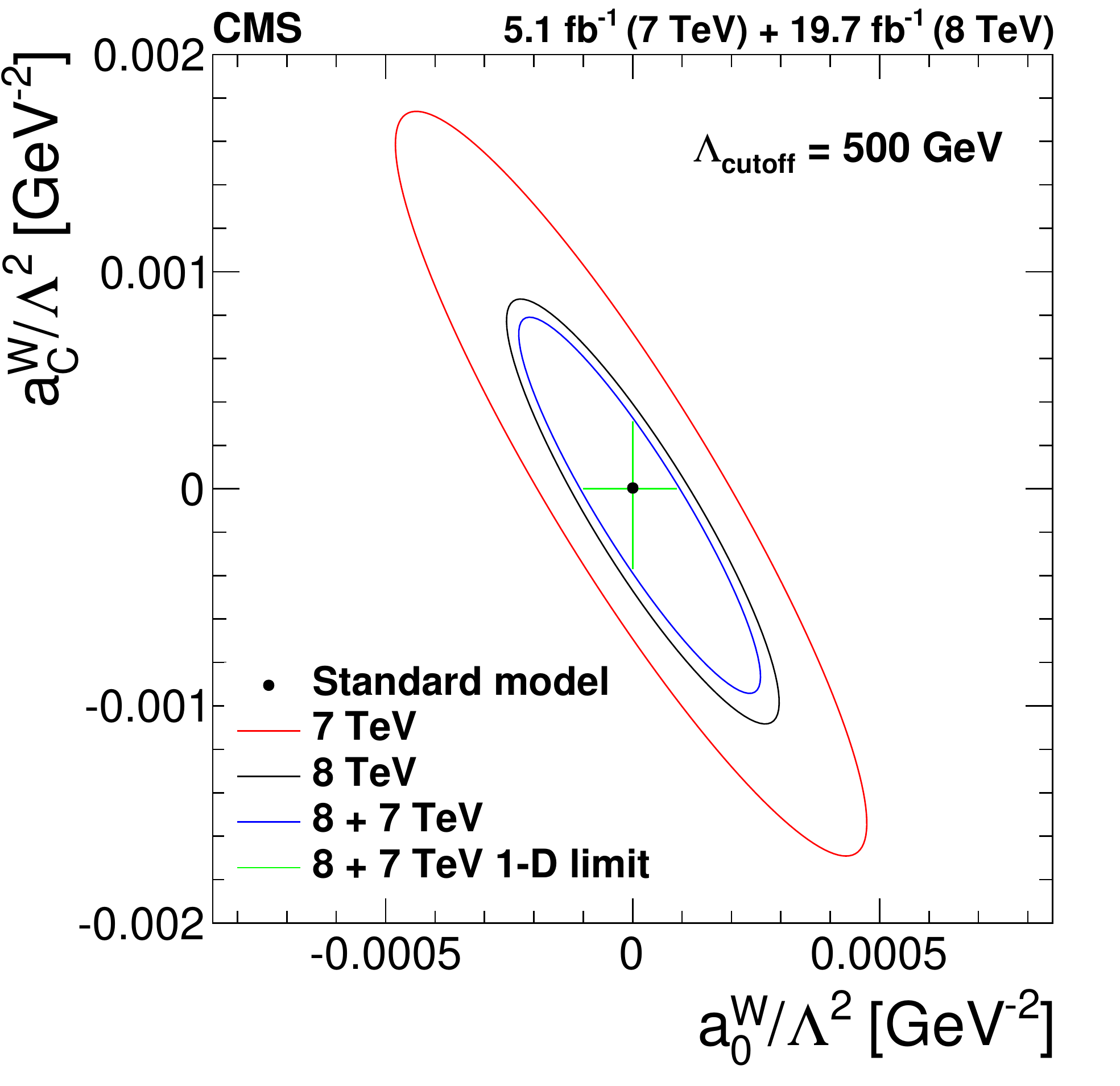}
\caption{Excluded values of the anomalous coupling parameters $a^{\PW}_{0}/\Lambda^{2}$ and $a^{\PW}_{C}/\Lambda^{2}$ with $\Lambda_{\text{cutoff}}=500\GeV$. The exclusion
regions are shown for the CMS measurements of $\PGg\PGg \to \PWp\PWm$ at 7\TeV (outer contour), 8\TeV (middle contour), and the 7+8\TeV combination (innermost contour).
The areas outside the solid contours are excluded by each measurement at 95\% CL. The cross indicates the one-dimensional limits obtained for each parameter from the 7 and 8\TeV
combination, with the other parameter fixed to zero.
\label{fig:Combo2DLimit}}
\end{figure}

We perform a similar procedure to derive two-dimensional limits in the ($a^{\PW}_{0}/\Lambda^{2}$, $a^{\PW}_{C}/\Lambda^{2}$) parameter space for the unitarized results
with $\Lambda_{\text{cutoff}}=500\GeV$.  The two-dimensional 95\% confidence level exclusion regions obtained from $\PGg\PGg \to \PWp\PWm$ production at CMS are shown
in Fig.~\ref{fig:Combo2DLimit} for the 7\TeV data (from Ref.~\cite{Chatrchyan:2013akv}), the 8\TeV data, and from the final 7 and 8\TeV combination.

\section{Conclusions}

Results are presented for exclusive and quasi-exclusive $\PGg\PGg \to \PWp\PWm$ production in the
$\PGm^{\pm}\Pe^{\mp}$ final state in pp collisions at $\sqrt{s} = 8\,(7)\TeV$, using data samples
corresponding to integrated luminosities of 19.7 (5.05)\fbinv.
In the signal region with $\pt(\PGm^{\pm}\Pe^{\mp}) > 30$\GeV and no
additional charged particles associated with the $\PGm^{\pm}\Pe^{\mp}$ vertex, we observe 13\,(2) events
with an expected background of $3.9 \pm 0.6$ ($0.84 \pm 0.15$) events in the 8 (7)\TeV data.
The observed yields and kinematic distributions are consistent with the SM prediction,
with a combined significance over the background-only
hypothesis of $3.4\sigma$. No significant deviations from the SM are observed in the $\pt(\PGm^{\pm}\Pe^{\mp})$
distribution, and the combined 7+8\TeV limits are interpreted in terms of improved constraints on dimension-6 and dimension-8
anomalous quartic gauge operator couplings.

\begin{acknowledgments}
\hyphenation{Bundes-ministerium Forschungs-gemeinschaft Forschungs-zentren} We congratulate our colleagues in the CERN accelerator departments for the excellent performance of the LHC and thank the technical and administrative staffs at CERN and at other CMS institutes for their contributions to the success of the CMS effort. In addition, we gratefully acknowledge the computing centers and personnel of the Worldwide LHC Computing Grid for delivering so effectively the computing infrastructure essential to our analyses. Finally, we acknowledge the enduring support for the construction and operation of the LHC and the CMS detector provided by the following funding agencies: the Austrian Federal Ministry of Science, Research and Economy and the Austrian Science Fund; the Belgian Fonds de la Recherche Scientifique, and Fonds voor Wetenschappelijk Onderzoek; the Brazilian Funding Agencies (CNPq, CAPES, FAPERJ, and FAPESP); the Bulgarian Ministry of Education and Science; CERN; the Chinese Academy of Sciences, Ministry of Science and Technology, and National Natural Science Foundation of China; the Colombian Funding Agency (COLCIENCIAS); the Croatian Ministry of Science, Education and Sport, and the Croatian Science Foundation; the Research Promotion Foundation, Cyprus; the Ministry of Education and Research, Estonian Research Council via IUT23-4 and IUT23-6 and European Regional Development Fund, Estonia; the Academy of Finland, Finnish Ministry of Education and Culture, and Helsinki Institute of Physics; the Institut National de Physique Nucl\'eaire et de Physique des Particules~/~CNRS, and Commissariat \`a l'\'Energie Atomique et aux \'Energies Alternatives~/~CEA, France; the Bundesministerium f\"ur Bildung und Forschung, Deutsche Forschungsgemeinschaft, and Helmholtz-Gemeinschaft Deutscher Forschungszentren, Germany; the General Secretariat for Research and Technology, Greece; the National Scientific Research Foundation, and National Innovation Office, Hungary; the Department of Atomic Energy and the Department of Science and Technology, India; the Institute for Studies in Theoretical Physics and Mathematics, Iran; the Science Foundation, Ireland; the Istituto Nazionale di Fisica Nucleare, Italy; the Ministry of Science, ICT and Future Planning, and National Research Foundation (NRF), Republic of Korea; the Lithuanian Academy of Sciences; the Ministry of Education, and University of Malaya (Malaysia); the Mexican Funding Agencies (CINVESTAV, CONACYT, SEP, and UASLP-FAI); the Ministry of Business, Innovation and Employment, New Zealand; the Pakistan Atomic Energy Commission; the Ministry of Science and Higher Education and the National Science Centre, Poland; the Funda\c{c}\~ao para a Ci\^encia e a Tecnologia, Portugal; JINR, Dubna; the Ministry of Education and Science of the Russian Federation, the Federal Agency of Atomic Energy of the Russian Federation, Russian Academy of Sciences, and the Russian Foundation for Basic Research; the Ministry of Education, Science and Technological Development of Serbia; the Secretar\'{\i}a de Estado de Investigaci\'on, Desarrollo e Innovaci\'on and Programa Consolider-Ingenio 2010, Spain; the Swiss Funding Agencies (ETH Board, ETH Zurich, PSI, SNF, UniZH, Canton Zurich, and SER); the Ministry of Science and Technology, Taipei; the Thailand Center of Excellence in Physics, the Institute for the Promotion of Teaching Science and Technology of Thailand, Special Task Force for Activating Research and the National Science and Technology Development Agency of Thailand; the Scientific and Technical Research Council of Turkey, and Turkish Atomic Energy Authority; the National Academy of Sciences of Ukraine, and State Fund for Fundamental Researches, Ukraine; the Science and Technology Facilities Council, UK; the US Department of Energy, and the US National Science Foundation.

Individuals have received support from the Marie-Curie program and the European Research Council and EPLANET (European Union); the Leventis Foundation; the A. P. Sloan Foundation; the Alexander von Humboldt Foundation; the Belgian Federal Science Policy Office; the Fonds pour la Formation \`a la Recherche dans l'Industrie et dans l'Agriculture (FRIA-Belgium); the Agentschap voor Innovatie door Wetenschap en Technologie (IWT-Belgium); the Ministry of Education, Youth and Sports (MEYS) of the Czech Republic; the Council of Science and Industrial Research, India; the HOMING PLUS program of the Foundation for Polish Science, cofinanced from European Union, Regional Development Fund; the Mobility Plus program of the Ministry of Science and Higher Education (Poland); the OPUS program of the National Science Center (Poland); MIUR project 20108T4XTM (Italy); the Thalis and Aristeia programs cofinanced by EU-ESF and the Greek NSRF; the National Priorities Research Program by Qatar National Research Fund; the Rachadapisek Sompot Fund for Postdoctoral Fellowship, Chulalongkorn University (Thailand); the Chulalongkorn Academic into Its 2nd Century Project Advancement Project (Thailand); and the Welch Foundation, contract C-1845.
\end{acknowledgments}

\bibliography{auto_generated}

\providecommand{\href}[2]{#2}\begingroup\raggedright\begin{thebibliography}{10}%
\makeatletter
\providecommand{\hrefCMSnoop }[0]{\@secondoftwo}%
\makeatother
\providecommand{\doi}{\texttt{doi:}\begingroup \urlstyle{tt}\Url}

\bibitem{deFavereaudeJeneret:2009db}
J.~de~Favereau~de Jeneret\hrefCMSnoop {}{ {et~al.}, ``High energy photon
  interactions at the {LHC}'',} (2009).
\href{http://www.arXiv.org/abs/0908.2020}{\texttt{arXiv:0908.2020}}.

\bibitem{Chatrchyan:2011ci}
\hrefCMSnoop {}{{CMS Collaboration}, ``{Exclusive photon-photon production of
  muon pairs in proton-proton collisions at $\sqrt{s}=7$ TeV}'',} \textit{
  JHEP} \textbf{ 01} (2012) 052,
  \href{http://dx.doi.org/10.1007/JHEP01(2012)052}{\doi{10.1007/JHEP01(2012)052}},
\href{http://www.arXiv.org/abs/1111.5536}{\texttt{arXiv:1111.5536}}.

\bibitem{Aad:2015bwa}
\hrefCMSnoop {}{{ATLAS Collaboration}, ``{Measurement of exclusive
  $\gamma\gamma\rightarrow \ell^+\ell^-$ production in proton-proton collisions
  at $\sqrt{s} = 7$ TeV with the ATLAS detector}'',} \textit{ Phys. Lett. B}
  \textbf{ 749} (2015) 242,
  \href{http://dx.doi.org/10.1016/j.physletb.2015.07.069}{\doi{10.1016/j.physletb.2015.07.069}},
\href{http://www.arXiv.org/abs/1506.07098}{\texttt{arXiv:1506.07098}}.

\bibitem{Chatrchyan:2012tv}
\hrefCMSnoop {}{{CMS Collaboration}, ``{Search for exclusive or semi-exclusive
  photon pair production and observation of exclusive and semi-exclusive
  electron pair production in $\mathrm{pp}$ collisions at $\sqrt{s}=7$ TeV}'',}
  \textit{ JHEP} \textbf{ 11} (2012) 080,
  \href{http://dx.doi.org/10.1007/JHEP11(2012)080}{\doi{10.1007/JHEP11(2012)080}},
\href{http://www.arXiv.org/abs/1209.1666}{\texttt{arXiv:1209.1666}}.

\bibitem{Chatrchyan:2013akv}
\hrefCMSnoop {}{{CMS Collaboration}, ``{Study of exclusive two-photon
  production of $W^+W^-$ in $\mathrm{pp}$ collisions at $\sqrt{s} = 7$ TeV and
  constraints on anomalous quartic gauge couplings}'',} \textit{ JHEP} \textbf{
  07} (2013) 116,
  \href{http://dx.doi.org/10.1007/JHEP07(2013)116}{\doi{10.1007/JHEP07(2013)116}},
\href{http://www.arXiv.org/abs/1305.5596}{\texttt{arXiv:1305.5596}}.

\bibitem{Pierzchala:2008xc}
\hrefCMSnoop {}{T.~Pierzchala and K.~Piotrzkowski, ``Sensitivity to anomalous
  quartic gauge couplings in photon-photon interactions at the {LHC}'',}
  \textit{ Nucl. Phys. Proc. Suppl.} \textbf{ 179} (2008) 257,
  \href{http://dx.doi.org/10.1016/j.nuclphysbps.2008.07.032}{\doi{10.1016/j.nuclphysbps.2008.07.032}},
\href{http://www.arXiv.org/abs/0807.1121}{\texttt{arXiv:0807.1121}}.

\bibitem{Chapon:2009hh}
\hrefCMSnoop {}{E.~Chapon, C.~Royon, and O.~Kepka, ``{Anomalous quartic $W W
  \gamma \gamma$, $Z Z \gamma \gamma$, and trilinear $WW \gamma$ couplings in
  two-photon processes at high luminosity at the LHC}'',} \textit{ Phys. Rev.
  D} \textbf{ 81} (2010) 074003,
  \href{http://dx.doi.org/10.1103/PhysRevD.81.074003}{\doi{10.1103/PhysRevD.81.074003}},
\href{http://www.arXiv.org/abs/0912.5161}{\texttt{arXiv:0912.5161}}.

\bibitem{Maniatis:2008zz}
\hrefCMSnoop {}{M.~Maniatis, A.~von Manteuffel, and O.~Nachtmann, ``{Anomalous
  couplings in $\gamma \gamma \to W^{+} W^{-}$ at LHC and ILC}'',} \textit{
  Nucl. Phys. Proc. Suppl.} \textbf{ 179} (2008) 104,
\href{http://dx.doi.org/10.1016/j.nuclphysbps.2008.07.012}{\doi{10.1016/j.nuclphysbps.2008.07.012}}.

\bibitem{Delgado:2014jda}
\hrefCMSnoop {}{R.~L. Delgado, A.~Dobado, M.~J. Herrero, and J.~J.
  Sanz-Cillero, ``{One-loop $\gamma\gamma \to W_{L}^{+} W_{L}^{-}$ and
  $\gamma\gamma \to Z_{L} Z_{L}$ from the electroweak chiral lagrangian with a
  light Higgs-like scalar}'',} \textit{ JHEP} \textbf{ 07} (2014) 149,
  \href{http://dx.doi.org/10.1007/JHEP07(2014)149}{\doi{10.1007/JHEP07(2014)149}},
\href{http://www.arXiv.org/abs/1404.2866}{\texttt{arXiv:1404.2866}}.

\bibitem{Fichet:2013ola}
\hrefCMSnoop {}{S.~Fichet and G.~von Gersdorff, ``{Anomalous gauge couplings
  from composite Higgs and warped extra dimensions}'',} \textit{ JHEP} \textbf{
  03} (2014) 102,
  \href{http://dx.doi.org/10.1007/JHEP03(2014)102}{\doi{10.1007/JHEP03(2014)102}},
\href{http://www.arXiv.org/abs/1311.6815}{\texttt{arXiv:1311.6815}}.

\bibitem{Espriu:2014jya}
\hrefCMSnoop {}{D.~Espriu and F.~Mescia, ``Unitarity and causality constraints
  in composite {H}iggs models'',} \textit{ Phys. Rev. D} \textbf{ 90} (2014)
  015035,
  \href{http://dx.doi.org/10.1103/PhysRevD.90.015035}{\doi{10.1103/PhysRevD.90.015035}},
\href{http://www.arXiv.org/abs/1403.7386}{\texttt{arXiv:1403.7386}}.

\bibitem{Heister:2004yd}
\hrefCMSnoop {}{{ALEPH} Collaboration, ``Constraints on anomalous QGC's in
  e$^{+}$e$^{-}$ interactions from 183~GeV to 209~GeV'',} \textit{ Phys. Lett.
  B} \textbf{ 602} (2004) 31,
\href{http://dx.doi.org/10.1016/j.physletb.2004.09.041}{\doi{10.1016/j.physletb.2004.09.041}}.

\bibitem{Abbiendi:2004bf}
\hrefCMSnoop {}{{OPAL} Collaboration, ``Constraints on anomalous quartic gauge
  boson couplings from $\nu\bar\nu\gamma\gamma$ and $q\bar q\gamma\gamma$
  events at LEP-2'',} \textit{ Phys. Rev. D} \textbf{ 70} (2004) 032005,
  \href{http://dx.doi.org/10.1103/PhysRevD.70.032005}{\doi{10.1103/PhysRevD.70.032005}},
\href{http://www.arXiv.org/abs/hep-ex/0402021}{\texttt{arXiv:hep-ex/0402021}}.

\bibitem{Abbiendi:2003jh}
\hrefCMSnoop {}{{OPAL} Collaboration, ``A study of $W^{+}W^{-}\gamma$ events at
  LEP'',} \textit{ Phys. Lett. B} \textbf{ 580} (2004) 17,
  \href{http://dx.doi.org/10.1016/j.physletb.2003.10.063}{\doi{10.1016/j.physletb.2003.10.063}},
\href{http://www.arXiv.org/abs/hep-ex/0309013}{\texttt{arXiv:hep-ex/0309013}}.

\bibitem{Abbiendi:1999aa}
\hrefCMSnoop {}{{OPAL} Collaboration, ``Measurement of the $W^{+}W^{-}\gamma$
  cross-section and first direct limits on anomalous electroweak quartic gauge
  couplings'',} \textit{ Phys. Lett. B} \textbf{ 471} (1999) 293,
  \href{http://dx.doi.org/10.1016/S0370-2693(99)01357-X}{\doi{10.1016/S0370-2693(99)01357-X}},
\href{http://www.arXiv.org/abs/hep-ex/9910069}{\texttt{arXiv:hep-ex/9910069}}.

\bibitem{Abdallah:2003xn}
\hrefCMSnoop {}{{DELPHI} Collaboration, ``Measurement of the e$^{+}$e$^{-}\to
  W^{+}W^{-}\gamma$ cross-section and limits on anomalous quartic gauge
  couplings with {DELPHI}'',} \textit{ Eur. Phys. J. C} \textbf{ 31} (2003)
  139,
  \href{http://dx.doi.org/10.1140/epjc/s2003-01350-x}{\doi{10.1140/epjc/s2003-01350-x}},
\href{http://www.arXiv.org/abs/hep-ex/0311004}{\texttt{arXiv:hep-ex/0311004}}.

\bibitem{Achard:2002iz}
\hrefCMSnoop {}{{L3} Collaboration, ``The e$^{+}$e$^{-}\to Z\gamma\gamma\to
  q\bar{q}\gamma \gamma$ reaction at {LEP} and constraints on anomalous quartic
  gauge boson couplings'',} \textit{ Phys. Lett. B} \textbf{ 540} (2002) 43,
  \href{http://dx.doi.org/10.1016/S0370-2693(02)02127-5}{\doi{10.1016/S0370-2693(02)02127-5}},
\href{http://www.arXiv.org/abs/hep-ex/0206050}{\texttt{arXiv:hep-ex/0206050}}.

\bibitem{Achard:2001eg}
\hrefCMSnoop {}{{L3} Collaboration, ``Study of the $W^{+}W^{-}\gamma$ process
  and limits on anomalous quartic gauge boson couplings at {LEP}'',} \textit{
  Phys. Lett. B} \textbf{ 527} (2002) 29,
  \href{http://dx.doi.org/10.1016/S0370-2693(02)01167-X}{\doi{10.1016/S0370-2693(02)01167-X}},
\href{http://www.arXiv.org/abs/hep-ex/0111029}{\texttt{arXiv:hep-ex/0111029}}.

\bibitem{Abazov:2013opa}
\hrefCMSnoop {}{{D0} Collaboration, ``{Search for anomalous quartic
  $WW\gamma\gamma$ couplings in dielectron and missing energy final states in
  p$\bar{\mathrm{p}}$ collisions at $\sqrt{s}$ = 1.96 TeV}'',} \textit{ Phys.
  Rev. D} \textbf{ 88} (2013) 012005,
  \href{http://dx.doi.org/10.1103/PhysRevD.88.012005}{\doi{10.1103/PhysRevD.88.012005}},
\href{http://www.arXiv.org/abs/1305.1258}{\texttt{arXiv:1305.1258}}.

\bibitem{Chatrchyan:2014bza}
\hrefCMSnoop {}{{CMS Collaboration}, ``{A search for WW gamma and WZ gamma
  production and constraints on anomalous quartic gauge couplings in
  $\mathrm{pp}$ collisions at sqrt(s) = 8 TeV}'',} \textit{ Phys. Rev. D}
  \textbf{ 90} (2014) 032008,
  \href{http://dx.doi.org/10.1103/PhysRevD.90.032008}{\doi{10.1103/PhysRevD.90.032008}},
\href{http://www.arXiv.org/abs/1404.4619}{\texttt{arXiv:1404.4619}}.

\bibitem{Aad:2015uqa}
\hrefCMSnoop {}{{ATLAS Collaboration}, ``{Evidence of $W\gamma \gamma$
  production in $\mathrm{pp}$ collisions at $\sqrt{s}=$ 8 TeV and limits on
  anomalous quartic gauge couplings with the ATLAS Detector}'',} \textit{ Phys.
  Rev. Lett.} \textbf{ 115} (2015) 031802,
  \href{http://dx.doi.org/10.1103/PhysRevLett.115.031802}{\doi{10.1103/PhysRevLett.115.031802}},
\href{http://www.arXiv.org/abs/1503.03243}{\texttt{arXiv:1503.03243}}.

\bibitem{Aad:2014zda}
\hrefCMSnoop {}{{ATLAS Collaboration}, ``{Evidence for electroweak production
  of $W^{\pm}W^{\pm}$jj in pp collisions at $\sqrt{s}=8$ TeV with the ATLAS
  detector}'',} \textit{ Phys. Rev. Lett.} \textbf{ 113} (2014) 141803,
  \href{http://dx.doi.org/10.1103/PhysRevLett.113.141803}{\doi{10.1103/PhysRevLett.113.141803}},
\href{http://www.arXiv.org/abs/1405.6241}{\texttt{arXiv:1405.6241}}.

\bibitem{Khachatryan:2014sta}
\hrefCMSnoop {}{{CMS Collaboration}, ``{Study of vector boson scattering and
  search for new physics in events with two same-sign leptons and two jets}'',}
  \textit{ Phys. Rev. Lett.} \textbf{ 114} (2015) 051801,
  \href{http://dx.doi.org/10.1103/PhysRevLett.114.051801}{\doi{10.1103/PhysRevLett.114.051801}},
\href{http://www.arXiv.org/abs/1410.6315}{\texttt{arXiv:1410.6315}}.

\bibitem{Khoze:2000db}
\hrefCMSnoop {}{V.~A. Khoze, A.~D. Martin, R.~Orava, and M.~G. Ryskin,
  ``{Luminosity monitors at the LHC}'',} \textit{ Eur. Phys. J. C} \textbf{ 19}
  (2001) 313,
  \href{http://dx.doi.org/10.1007/s100520100616}{\doi{10.1007/s100520100616}},
\href{http://www.arXiv.org/abs/hep-ph/0010163}{\texttt{arXiv:hep-ph/0010163}}.

\bibitem{HarlandLang:2012qz}
\hrefCMSnoop {}{L.~A. Harland-Lang, V.~A. Khoze, M.~G. Ryskin, and W.~J.
  Stirling, ``The phenomenology of central exclusive production at hadron
  colliders'',} \textit{ Eur. Phys. J. C} \textbf{ 72} (2012) 2110,
  \href{http://dx.doi.org/10.1140/epjc/s10052-012-2110-2}{\doi{10.1140/epjc/s10052-012-2110-2}},
\href{http://www.arXiv.org/abs/1204.4803}{\texttt{arXiv:1204.4803}}.

\bibitem{Belanger:1992qh}
\hrefCMSnoop {}{G.~Belanger and F.~Boudjema, ``Probing quartic couplings of
  weak bosons through three vectors production at a {500 GeV NLC}'',} \textit{
  Phys. Lett. B} \textbf{ 288} (1992) 201,
\href{http://dx.doi.org/10.1016/0370-2693(92)91978-I}{\doi{10.1016/0370-2693(92)91978-I}}.

\bibitem{Aad:2012tfa}
\hrefCMSnoop {}{{ATLAS Collaboration}, ``{Observation of a new particle in the
  search for the Standard Model Higgs boson with the ATLAS detector at the
  LHC}'',} \textit{ Phys. Lett. B} \textbf{ 716} (2012) 1,
  \href{http://dx.doi.org/10.1016/j.physletb.2012.08.020}{\doi{10.1016/j.physletb.2012.08.020}},
\href{http://www.arXiv.org/abs/1207.7214}{\texttt{arXiv:1207.7214}}.

\bibitem{Chatrchyan:2012ufa}
\hrefCMSnoop {}{{CMS Collaboration}, ``{Observation of a new boson at a mass of
  125 GeV with the CMS experiment at the LHC}'',} \textit{ Phys. Lett. B}
  \textbf{ 716} (2012) 30,
  \href{http://dx.doi.org/10.1016/j.physletb.2012.08.021}{\doi{10.1016/j.physletb.2012.08.021}},
\href{http://www.arXiv.org/abs/1207.7235}{\texttt{arXiv:1207.7235}}.

\bibitem{Chatrchyan:2013lba}
\hrefCMSnoop {}{{CMS Collaboration}, ``{Observation of a new boson with mass
  near 125 GeV in pp collisions at $\sqrt{s}$ = 7 and 8 TeV}'',} \textit{ JHEP}
  \textbf{ 06} (2013) 081,
  \href{http://dx.doi.org/10.1007/JHEP06(2013)081}{\doi{10.1007/JHEP06(2013)081}},
\href{http://www.arXiv.org/abs/1303.4571}{\texttt{arXiv:1303.4571}}.

\bibitem{Belanger:1999aw}
G.~Belanger\hrefCMSnoop {}{ {et~al.}, ``{Bosonic quartic couplings at
  LEP-2}'',} \textit{ Eur. Phys. J. C} \textbf{ 13} (2000) 283,
  \href{http://dx.doi.org/10.1007/s100520000305}{\doi{10.1007/s100520000305}},
\href{http://www.arXiv.org/abs/hep-ph/9908254}{\texttt{arXiv:hep-ph/9908254}}.

\bibitem{Eboli:2006wa}
\hrefCMSnoop {}{O.~J.~P. Eboli, M.~C. Gonzalez-Garcia, and J.~K. Mizukoshi,
  ``{$p p \rightarrow$jje$^{\pm} \mu^{\pm} \nu \nu$ and jje$^{\pm} \mu^{\mp}
  \nu \nu$ at O($\alpha_{em}^{6}$) and O($\alpha_{em}^{4} \alpha_{s}^{2}$) for
  the study of the quartic electroweak gauge boson vertex at CERN LHC}'',}
  \textit{ Phys. Rev. D} \textbf{ 74} (2006) 073005,
  \href{http://dx.doi.org/10.1103/PhysRevD.74.073005}{\doi{10.1103/PhysRevD.74.073005}},
\href{http://www.arXiv.org/abs/hep-ph/0606118}{\texttt{arXiv:hep-ph/0606118}}.

\bibitem{Baak:2013fwa}
\hrefCMSnoop {}{M.~Baak {et~al.}, ``{Working group report: precision study of
  electroweak interactions}'',} (2013).
\href{http://www.arXiv.org/abs/1310.6708}{\texttt{arXiv:1310.6708}}.

\bibitem{Eboli:2000ad}
\hrefCMSnoop {}{O.~J.~P. Eboli, M.~C. Gonzalez-Garcia, S.~M. Lietti, and S.~F.
  Novaes, ``Anomalous quartic gauge boson couplings at hadron colliders'',}
  \textit{ Phys. Rev. D} \textbf{ 63} (2001) 075008,
  \href{http://dx.doi.org/10.1103/PhysRevD.63.075008}{\doi{10.1103/PhysRevD.63.075008}},
\href{http://www.arXiv.org/abs/hep-ph/0009262}{\texttt{arXiv:hep-ph/0009262}}.

\bibitem{Gupta:2011be}
\hrefCMSnoop {}{R.~S. Gupta, ``{Probing quartic neutral gauge boson couplings
  using diffractive photon fusion at the LHC}'',} \textit{ Phys. Rev. D}
  \textbf{ 85} (2012) 014006,
  \href{http://dx.doi.org/10.1103/PhysRevD.85.014006}{\doi{10.1103/PhysRevD.85.014006}},
\href{http://www.arXiv.org/abs/1111.3354}{\texttt{arXiv:1111.3354}}.

\bibitem{Alboteanu:2008my}
\hrefCMSnoop {}{A.~Alboteanu, W.~Kilian, and J.~Reuter, ``{Resonances and
  unitarity in weak boson scattering at the LHC}'',} \textit{ JHEP} \textbf{
  11} (2008) 010,
  \href{http://dx.doi.org/10.1088/1126-6708/2008/11/010}{\doi{10.1088/1126-6708/2008/11/010}},
\href{http://www.arXiv.org/abs/0806.4145}{\texttt{arXiv:0806.4145}}.

\bibitem{JINST}
\hrefCMSnoop {}{{CMS Collaboration}, ``The {CMS} experiment at the {CERN}
  {LHC}'',} \textit{ JINST} \textbf{ 3} (2008) S08004,
\href{http://dx.doi.org/10.1088/1748-0221/3/08/S08004}{\doi{10.1088/1748-0221/3/08/S08004}}.

\bibitem{Khachatryan:2015hwa}
\hrefCMSnoop {}{{CMS Collaboration}, ``{Performance of electron reconstruction
  and selection with the CMS detector in proton-proton collisions at $\sqrt{s}=
  $ 8 TeV}'',} \textit{ JINST} \textbf{ 10} (2015) P06005,
  \href{http://dx.doi.org/10.1088/1748-0221/10/06/P06005}{\doi{10.1088/1748-0221/10/06/P06005}},
\href{http://www.arXiv.org/abs/1502.02701}{\texttt{arXiv:1502.02701}}.

\bibitem{Chatrchyan:2014fea}
\hrefCMSnoop {}{{CMS Collaboration}, ``{Description and performance of track
  and primary-vertex reconstruction with the CMS tracker}'',} \textit{ JINST}
  \textbf{ 9} (2014) P10009,
  \href{http://dx.doi.org/10.1088/1748-0221/9/10/P10009}{\doi{10.1088/1748-0221/9/10/P10009}},
\href{http://www.arXiv.org/abs/1405.6569}{\texttt{arXiv:1405.6569}}.

\bibitem{Alwall:2011uj}
J.~Alwall\hrefCMSnoop {}{ {et~al.}, ``{MadGraph 5: going beyond}'',} \textit{
  JHEP} \textbf{ 06} (2011) 128,
  \href{http://dx.doi.org/10.1007/JHEP06(2011)128}{\doi{10.1007/JHEP06(2011)128}},
\href{http://www.arXiv.org/abs/1106.0522}{\texttt{arXiv:1106.0522}}.

\bibitem{Alwall:2014hca}
J.~Alwall\hrefCMSnoop {}{ {et~al.}, ``{The automated computation of tree-level
  and next-to-leading order differential cross sections, and their matching to
  parton shower simulations}'',} \textit{ JHEP} \textbf{ 07} (2014) 079,
  \href{http://dx.doi.org/10.1007/JHEP07(2014)079}{\doi{10.1007/JHEP07(2014)079}},
\href{http://www.arXiv.org/abs/1405.0301}{\texttt{arXiv:1405.0301}}.

\bibitem{Budnev:1974de}
\hrefCMSnoop {}{V.~M. Budnev, I.~F. Ginzburg, G.~V. Meledin, and V.~G. Serbo,
  ``{The two-photon particle production mechanism. Physical problems.
  Applications. Equivalent photon approximation}'',} \textit{ Phys. Rept.}
  \textbf{ 15} (1975) 181,
\href{http://dx.doi.org/10.1016/0370-1573(75)90009-5}{\doi{10.1016/0370-1573(75)90009-5}}.

\bibitem{Belyaev:2012qa}
\hrefCMSnoop {}{A.~Belyaev, N.~D. Christensen, and A.~Pukhov, ``{CalcHEP 3.4
  for collider physics within and beyond the Standard Model}'',} \textit{
  Comput. Phys. Commun.} \textbf{ 184} (2013) 1729,
  \href{http://dx.doi.org/10.1016/j.cpc.2013.01.014}{\doi{10.1016/j.cpc.2013.01.014}},
\href{http://www.arXiv.org/abs/1207.6082}{\texttt{arXiv:1207.6082}}.

\bibitem{Vermaseren:1982cz}
\hrefCMSnoop {}{J.~A.~M. Vermaseren, ``Two-photon processes at very high
  energies'',} \textit{ Nucl. Phys. B} \textbf{ 229} (1983) 347,
\href{http://dx.doi.org/10.1016/0550-3213(83)90336-X}{\doi{10.1016/0550-3213(83)90336-X}}.

\bibitem{Baranov:1991yq}
\hrefCMSnoop {}{S.~P. Baranov, O.~Duenger, H.~Shooshtari, and J.~A.~M.
  Vermaseren, ``{LPAIR}: A generator for lepton pair production'',} in \textit{
  Proceedings of Physics at HERA}, volume~3, p.~1478.
\newblock
1991.
\newblock

\bibitem{Czakon:2011xx}
\hrefCMSnoop {}{M.~Czakon and A.~Mitov, ``{Top++: a program for the calculation
  of the top-pair cross-section at hadron colliders}'',} \textit{ Comput. Phys.
  Commun.} \textbf{ 185} (2014) 2930,
  \href{http://dx.doi.org/10.1016/j.cpc.2014.06.021}{\doi{10.1016/j.cpc.2014.06.021}},
\href{http://www.arXiv.org/abs/1112.5675}{\texttt{arXiv:1112.5675}}.

\bibitem{Campbell:2010ff}
\hrefCMSnoop {}{J.~M. Campbell and R.~K. Ellis, ``{MCFM for the Tevatron and
  the LHC}'',} \textit{ Nucl. Phys. Proc. Suppl.} \textbf{ 205} (2010) 10,
  \href{http://dx.doi.org/10.1016/j.nuclphysbps.2010.08.011}{\doi{10.1016/j.nuclphysbps.2010.08.011}},
\href{http://www.arXiv.org/abs/1007.3492}{\texttt{arXiv:1007.3492}}.

\bibitem{Nason:2004rx}
\hrefCMSnoop {}{P.~Nason, ``A new method for combining {NLO} {QCD} with shower
  Monte Carlo algorithms'',} \textit{ JHEP} \textbf{ 11} (2004) 040,
  \href{http://dx.doi.org/10.1088/1126-6708/2004/11/040}{\doi{10.1088/1126-6708/2004/11/040}},
\href{http://www.arXiv.org/abs/hep-ph/0409146}{\texttt{arXiv:hep-ph/0409146}}.

\bibitem{Frixione:2007vw}
\hrefCMSnoop {}{S.~Frixione, P.~Nason, and C.~Oleari, ``Matching {NLO} {QCD}
  computations with parton shower simulations: the {POWHEG} method'',} \textit{
  JHEP} \textbf{ 11} (2007) 070,
  \href{http://dx.doi.org/10.1088/1126-6708/2007/11/070}{\doi{10.1088/1126-6708/2007/11/070}},
\href{http://www.arXiv.org/abs/0709.2092}{\texttt{arXiv:0709.2092}}.

\bibitem{Alioli:2010xd}
\hrefCMSnoop {}{S.~Alioli, P.~Nason, C.~Oleari, and E.~Re, ``A general
  framework for implementing {NLO} calculations in shower Monte Carlo programs:
  the {POWHEG BOX}'',} \textit{ JHEP} \textbf{ 06} (2010) 043,
  \href{http://dx.doi.org/10.1007/JHEP06(2010)043}{\doi{10.1007/JHEP06(2010)043}},
\href{http://www.arXiv.org/abs/1002.2581}{\texttt{arXiv:1002.2581}}.

\bibitem{Sjostrand:2006za}
\hrefCMSnoop {}{T.~Sj{\"o}strand, S.~Mrenna, and P.~Z. Skands, ``{PYTHIA 6.4
  Physics and Manual}'',} \textit{ JHEP} \textbf{ 05} (2006) 026,
  \href{http://dx.doi.org/10.1088/1126-6708/2006/05/026}{\doi{10.1088/1126-6708/2006/05/026}},
\href{http://www.arXiv.org/abs/hep-ph/0603175}{\texttt{arXiv:hep-ph/0603175}}.

\bibitem{Chatrchyan:2013gfi}
\hrefCMSnoop {}{{CMS Collaboration}, ``{Study of the underlying event at
  forward rapidity in pp collisions at $\sqrt{s}$ = 0.9, 2.76, and 7 TeV}'',}
  \textit{ JHEP} \textbf{ 04} (2013) 072,
  \href{http://dx.doi.org/10.1007/JHEP04(2013)072}{\doi{10.1007/JHEP04(2013)072}},
\href{http://www.arXiv.org/abs/1302.2394}{\texttt{arXiv:1302.2394}}.

\bibitem{Field2011}
\hrefCMSnoop {}{R.~Field, ``Min-bias and the underlying event at the {LHC}'',}
  \textit{ Acta Phys. Polon. B} \textbf{ 42} (2011) 2631,
  \href{http://dx.doi.org/10.5506/APhysPolB.42.2631}{\doi{10.5506/APhysPolB.42.2631}},
\href{http://www.arXiv.org/abs/1110.5530}{\texttt{arXiv:1110.5530}}.

\bibitem{Bruni:1993is}
\hrefCMSnoop {}{P.~Bruni and G.~Ingelman, ``Diffractive hard scattering at
  $\mathrm{ep}$ and p$\bar{\mathrm{p}}$ colliders'',} in \textit{ Proceedings
  of the Int. Europhysics Conf. on High Energy Physics}, J.~Carr and
  M.~Perrottet, eds., p.~595.
\newblock Ed. Fronti{\`e}res, Marseille, France,
1994.
\newblock

\bibitem{Aktas:2006hy}
\hrefCMSnoop {}{{H1} Collaboration, ``{Measurement and QCD analysis of the
  diffractive deep-inelastic scattering cross-section at HERA}'',} \textit{
  Eur. Phys. J. C} \textbf{ 48} (2006) 715,
  \href{http://dx.doi.org/10.1140/epjc/s10052-006-0035-3}{\doi{10.1140/epjc/s10052-006-0035-3}},
\href{http://www.arXiv.org/abs/hep-ex/0606004}{\texttt{arXiv:hep-ex/0606004}}.

\bibitem{Chatrchyan:2012vc}
\hrefCMSnoop {}{{CMS Collaboration}, ``{Observation of a diffractive
  contribution to dijet production in proton-proton collisions at $\sqrt{s}=7$
  TeV}'',} \textit{ Phys. Rev. D} \textbf{ 87} (2013) 012006,
  \href{http://dx.doi.org/10.1103/PhysRevD.87.012006}{\doi{10.1103/PhysRevD.87.012006}},
\href{http://www.arXiv.org/abs/1209.1805}{\texttt{arXiv:1209.1805}}.

\bibitem{Lebiedowicz:2012gg}
\hrefCMSnoop {}{P.~Lebiedowicz, R.~Pasechnik, and A.~Szczurek, ``{QCD
  diffractive mechanism of exclusive $W^+W^-$ pair production at high
  energies}'',} \textit{ Nucl. Phys. B} \textbf{ 867} (2013) 61,
  \href{http://dx.doi.org/10.1016/j.nuclphysb.2012.09.014}{\doi{10.1016/j.nuclphysb.2012.09.014}},
\href{http://www.arXiv.org/abs/1203.1832}{\texttt{arXiv:1203.1832}}.

\bibitem{Khachatryan:2014dea}
\hrefCMSnoop {}{{CMS Collaboration}, ``{Measurement of electroweak production
  of two jets in association with a Z boson in proton-proton collisions at
  $\sqrt{s}=8\,\text {TeV}$}'',} \textit{ Eur. Phys. J. C} \textbf{ 75} (2015)
  66,
  \href{http://dx.doi.org/10.1140/epjc/s10052-014-3232-5}{\doi{10.1140/epjc/s10052-014-3232-5}},
\href{http://www.arXiv.org/abs/1410.3153}{\texttt{arXiv:1410.3153}}.

\bibitem{Ballestrero:2007xq}
A.~Ballestrero\hrefCMSnoop {}{ {et~al.}, ``{PHANTOM: A Monte Carlo event
  generator for six parton final states at high energy colliders}'',} \textit{
  Comput. Phys. Commun.} \textbf{ 180} (2009) 401,
  \href{http://dx.doi.org/10.1016/j.cpc.2008.10.005}{\doi{10.1016/j.cpc.2008.10.005}},
\href{http://www.arXiv.org/abs/0801.3359}{\texttt{arXiv:0801.3359}}.

\bibitem{Agostinelli:2002hh}
\hrefCMSnoop {}{{GEANT4} Collaboration, ``{GEANT4}---a simulation toolkit'',}
  \textit{ Nucl. Instrum. Meth. A} \textbf{ 506} (2003) 250,
\href{http://dx.doi.org/10.1016/S0168-9002(03)01368-8}{\doi{10.1016/S0168-9002(03)01368-8}}.

\bibitem{Chatrchyan:2012xi}
\hrefCMSnoop {}{{CMS Collaboration}, ``{Performance of CMS muon reconstruction
  in pp collision events at $\sqrt{s}=7$~TeV}'',} \textit{ JINST} \textbf{ 7}
  (2012) P10002,
  \href{http://dx.doi.org/10.1088/1748-0221/7/10/P10002}{\doi{10.1088/1748-0221/7/10/P10002}},
\href{http://www.arXiv.org/abs/1206.4071}{\texttt{arXiv:1206.4071}}.

\bibitem{CMS:2009nxa}
\href {http://cdsweb.cern.ch/record/1194487}{{CMS Collaboration},
  ``Particle--Flow Event Reconstruction in {CMS} and Performance for Jets,
  Taus, and {\MET}'',} CMS Physics Analysis Summary CMS-PAS-PFT-09-001, 2009.

\bibitem{CMS:2010nxa}
\href {http://cdsweb.cern.ch/record/1247373}{{CMS Collaboration},
  ``Commissioning of the Particle-flow Event Reconstruction with the first
  {LHC} collisions recorded in the {CMS} detector'',} CMS Physics Analysis
  Summary CMS-PAS-PFT-10-001, 2010.

\bibitem{daSilveira:2014jla}
G.~G. da~Silveira\hrefCMSnoop {}{ {et~al.}, ``{Central $\mu^{+}$ $\mu^{-}$
  production via photon-photon fusion in proton-proton collisions with proton
  dissociation}'',} \textit{ JHEP} \textbf{ 02} (2015) 159,
  \href{http://dx.doi.org/10.1007/JHEP02(2015)159}{\doi{10.1007/JHEP02(2015)159}},
\href{http://www.arXiv.org/abs/1409.1541}{\texttt{arXiv:1409.1541}}.

\bibitem{Harland-Lang:2016apc}
\hrefCMSnoop {}{L.~A. Harland-Lang, V.~A. Khoze, and M.~G. Ryskin, ``{The
  photon PDF in events with rapidity gaps}'',} \textit{ Eur. Phys. J.} \textbf{
  C76} (2016), no.~5, 255,
  \href{http://dx.doi.org/10.1140/epjc/s10052-016-4100-2}{\doi{10.1140/epjc/s10052-016-4100-2}},
\href{http://www.arXiv.org/abs/1601.03772}{\texttt{arXiv:1601.03772}}.

\bibitem{CMS-PAS-LUM-13-001}
\href {http://cdsweb.cern.ch/record/1598864}{{CMS Collaboration}, ``CMS
  Luminosity Based on Pixel Cluster Counting - Summer 2013 Update'',} CMS
  Physics Analysis Summary CMS-PAS-LUM-13-001, 2013.

\bibitem{Feldman:1997qc}
\hrefCMSnoop {}{G.~J. Feldman and R.~D. Cousins, ``{A unified approach to the
  classical statistical analysis of small signals}'',} \textit{ Phys. Rev. D}
  \textbf{ 57} (1998) 3873,
  \href{http://dx.doi.org/10.1103/PhysRevD.57.3873}{\doi{10.1103/PhysRevD.57.3873}},
\href{http://www.arXiv.org/abs/physics/9711021}{\texttt{arXiv:physics/9711021}}.

\bibitem{Cousins:1991qz}
\hrefCMSnoop {}{R.~D. Cousins and V.~L. Highland, ``{Incorporating systematic
  uncertainties into an upper limit}'',} \textit{ Nucl. Instrum. Meth. A}
  \textbf{ 320} (1992) 331,
\href{http://dx.doi.org/10.1016/0168-9002(92)90794-5}{\doi{10.1016/0168-9002(92)90794-5}}.

\end{thebibliography}\endgroup

\cleardoublepage \appendix\section{The CMS Collaboration \label{app:collab}}\begin{sloppypar}\hyphenpenalty=5000\widowpenalty=500\clubpenalty=5000\textbf{Yerevan Physics Institute,  Yerevan,  Armenia}\\*[0pt]
V.~Khachatryan, A.M.~Sirunyan, A.~Tumasyan
\vskip\cmsinstskip
\textbf{Institut f\"{u}r Hochenergiephysik der OeAW,  Wien,  Austria}\\*[0pt]
W.~Adam, E.~Asilar, T.~Bergauer, J.~Brandstetter, E.~Brondolin, M.~Dragicevic, J.~Er\"{o}, M.~Flechl, M.~Friedl, R.~Fr\"{u}hwirth\cmsAuthorMark{1}, V.M.~Ghete, C.~Hartl, N.~H\"{o}rmann, J.~Hrubec, M.~Jeitler\cmsAuthorMark{1}, V.~Kn\"{u}nz, A.~K\"{o}nig, M.~Krammer\cmsAuthorMark{1}, I.~Kr\"{a}tschmer, D.~Liko, T.~Matsushita, I.~Mikulec, D.~Rabady\cmsAuthorMark{2}, N.~Rad, B.~Rahbaran, H.~Rohringer, J.~Schieck\cmsAuthorMark{1}, R.~Sch\"{o}fbeck, J.~Strauss, W.~Treberer-Treberspurg, W.~Waltenberger, C.-E.~Wulz\cmsAuthorMark{1}
% ...
\vskip\cmsinstskip
% ...
1:~~Also at Vienna University of Technology, Vienna, Austria\\
2:~~Also at CERN, European Organization for Nuclear Research, Geneva, Switzerland\\
% ...

\end{sloppypar}
\end{document}